\documentclass[a4paper, journal, 10pt]{IEEEtran}
\usepackage[T1]{fontenc}
\usepackage[utf8]{inputenc}
\usepackage{graphicx}
\usepackage{amsmath}
\usepackage{cleveref}
\usepackage{lipsum}
\usepackage{subfiles}
\usepackage{multirow}
\usepackage{subfig}
\usepackage{multirow}
\usepackage{url}

\usepackage{pifont}
\usepackage{xcolor}
\definecolor{DarkGreen}{rgb}{0.01,0.75,0.24}
\definecolor{DarkRed}{rgb}{0.8,0.31,0.36}
\newcommand{\cmark}{\ding{51}}%
\newcommand{\xmark}{\ding{55}}%
\usepackage{array}
\newcolumntype{x}[1]{>{\centering\arraybackslash\hspace{0pt}}p{#1}}
\usepackage{adjustbox}
\newcolumntype{R}[2]{%
    >{\adjustbox{angle=#1,lap=\width-(#2)}\bgroup}%
    l%
    <{\egroup}%
}
\newcommand*\rot{\multicolumn{1}{R{45}{1em}}}%

\newcommand{\STAB}[1]{\begin{tabular}{@{}c@{}}#1\end{tabular}}

\usepackage{eurosym}
\DeclareUnicodeCharacter{20AC}{\euro}

\usepackage[numbers]{natbib}
\title{\vspace{-0.5em}Web3: A Decentralized Societal Infrastructure for Identity, Trust, Money, and Data}
\author{\textbf{J.W. Bambacht and J.A. Pouwelse} \\ J.W.Bambacht@student.tudelft.nl, J.A.Pouwelse@tudelft.nl \\ Distributed Systems, Delft University of Technology \\ \today \vspace{-2.25em}}

\begin{document}
\maketitle

\begin{abstract}
    A movement for a more transparent and decentralized Internet is globally attracting more attention. People are becoming more privacy-aware of their online identities and data. The Internet is constantly evolving. Web2 focused on companies that provide services in exchange for personal user data. Web3 commits to user-centricity using decentralization and zero-server architectures. The current digital society demands a global change to empower citizens and take back control. Citizens are locked into big-tech for personal data storage and their for-profit digital identity. Protection of data has proven to be essential, especially due to increased home Internet traffic during the COVID pandemic. Citizens do not possess their own travel documents. The European Commission aims to transition this governmental property towards self-sovereign identity, introducing many new opportunities. Citizens are locked into banks with non-portable IBAN accounts and unsustainable legacy banking infrastructures. Migration to all-digital low-fraud infrastructures and healthier competitive ecosystems is essential.
    The overall challenge is to return the power to citizens and users again. The transition to a more decentralized Internet is the first crucial step in the realization of user-centricity. This thesis presents the first exploratory study that integrates governmental-issued travel documents into a (decentralized) societal infrastructure. These self-sovereign identities form the authentic base to a private and secure transfer of money and data, and can effectively provide trust in authenticity that is currently missing in online conversations. A fully operational zero-server infrastructure that incorporates all our requirements has been developed for Android using the P2P network overlay IPv8 \cite{ipv8-2020}, and a personalized blockchain called TrustChain \cite{trustchain-2020}. It contributes to a reformed tech and financial sector that is more efficient and effective in serving the wider economy, and more resistant to bad behavior of all kinds. Creating such an infrastructure that is decentralized and anti-fragile is deemed crucial for the future.
\end{abstract} 

\vspace{-0.75em}

\section{Introduction}
The online world is dominated by big-tech monopolies. These companies hold a relatively large amount of power in relation to citizens. As a result, citizens have difficulty protecting their data. The WhatsApp messaging platform is a motivating example of market failure as it violates terms of service over a long period \cite{whatsapptos-zingales2017}. An update of their terms of service \cite{whatsapptos-2021}, providing mother company Facebook access to more user data, initiated a migration to other platforms. Competitors focused on privacy and openness have barriers to market entry, no network effect, and compete against long existent closed protocols. Citizens and small(er) competitors \cite{big-tech-regulations2022} are powerless in this uncompetitive market. 

Digitization generally weakens the privacy of citizens. The European Union started an ongoing effort into the General Data Protection Regulation (GDPR) \cite{gdpr-2016} in 2016, targeting the misuse of privacy-sensitive data by companies. Many companies and platforms failed to comply to personal data protection, resulting in over 900 filed cases of GDPR complaints, with a total value over 1.3 billion euros \cite{gdprtracker}. Due to such efforts, citizens have become more privacy-aware of their online data and identities \cite{privacy-aikaterini2021}. Commercial entities have an incentive to minimize spending on cybersecurity \cite{data-breach2021}. The storage of personal data and weak security mechanisms of platforms are both at the expense of the user. These companies often gain revenue and value by selling user data for personalized advertisements. Users have no other option but to rely on the best intentions of the platform owner on their data. 

In many situations, personally identifiable information of citizens is unnecessarily exposed. Government-issued documents are often required for institutions or organizations like banks, insurance companies, hotels, and employers, but lack secure handling and storage. Online governmental authentication mechanisms for digital identities are widely deployed by authorized institutions but have equal concerns. The owners of these identities are forced to accept and transfer all personal information. Both citizens and governments could benefit from the use of self-sovereign identities (SSI). As a result, citizens will be given the control over their own identities. Furthermore, external dependencies on authentication and storage can be eliminated. Offline identity authentication and identity attestations in a user-centric fashion can still offer the required identity authenticity. The possibilities of the SSI are additionally suited to a wide range of applications. We can profit from the self-sovereign identity within our societal infrastructure for the enforcement of authentic trust between identities in online conversations with the goal to reduce phishing or impersonating attacks. Other applications may include authenticated signing of digital documents, and validated storage of diplomas and COVID vaccination certificates.

Governments, banks, and tax offices have general insight into the bank accounts and transactions of citizens, often using big-tech cloud services. Not only is this a violation of citizens' privacy, but these banking services also have several deficits. The transaction costs are disproportional as debit card transactions range from \euro0,05 to \euro0,20 per transaction \cite{offline-transaction-costs}, and even bigger numbers for external payment services like iDEAL \cite{ideal-costs2021}. Cross-border payments even have increased costs and settlements. Cash is only applicable in offline payments and still serves as a store of value for some. It offers respectable privacy but is (slowly) fading away \cite{factsheet-payments2020}. The current financial system can benefit from the adoption of blockchain technology. Central Bank Digital Currencies (CBDC) aims to provide a faster, more efficient, and cheaper alternative to electronic payments. With characteristics as privacy-awareness, pseudo-anonymity, and unrestricted cross-border payments, CBDC can transform the financial system into a sustainable and frictionless system. \\

This research contributes the following: (I) design of a novel decentralized infrastructure that incorporates a self-sovereign identity, (II) applicance of the self-sovereign identity for identity attestations and authentic trust enforcement between participants of communication, (III) generic transfer of money, (IV) generic transfer of data using a custom-designed P2P data transfer protocol.
\section{Problem Description}
The goal of this study is to design a novel decentralized societal infrastructure that incorporates a self-sovereign identity as an authentic base. We can additionally apply this identity in a useful way to facilitate authentic trust between users, and to privately transfer money and data with other identities. In a centralized infrastructure, platform owners are still able to collect metadata belonging to encrypted data. Removing these single points of failure reduces the violation of privacy and security of users.

The key aspect of our research is the application of citizens' self-sovereign identities. Self-sovereign identity is characterized by the ten principles of Allen \cite{ssiproperties-allen2016, ssideployment-stokking2018}, that try to assure the users' control within its own SSI, with a balance between transparency, fairness, and protection. Governments currently have ownership and control of these identities. Various opportunities arise by moving the power back to the user. Firstly, the user is the owner of their own identity and can view and decide what information to share. Secondly, as governments don't have control anymore, less personal data management is required, less bureaucracy, and a cost reduction for facilitating the heavily secured infrastructure and authentication mechanisms. Thirdly, the decrease of personal data on central servers reduces the possibility of data breaches and theft. And finally an opportunity arises to replace visual identity document checks. 
Currently, identity documents are exposed upon request of some authority. Not only does the requested information become visible, but also the complete document. By using identity attestations, authorities are able to verify information of the identity without exposing the actual value, e.g. age validation of a bouncer in the pub. Communication channels lack trust in the authenticity of other participants' online identities. Since we have access to an official SSI, we can even apply this information to build trust. No social platform currently integrates government-issued identity information in such a fashion. Generally, online identities consist of a name, picture, phone number, or email, that altogether contribute to confidence in the authenticity of the other user's identity. However, these attributes are manually forgeable and can be impersonated. Malicious actors try to make their fake identities look as genuine as possible to mimic someone's identity. Without these editable components and with automatic acquiring of information from the SSI, we can enforce authentic trust to other participants.

The global financial infrastructure is failing to provide real-time cross-border payments and is dominated by monopoly players with high profits and near-zero innovation. Financial privacy disappeared in the last decade(s). Governments, banks, and tax offices heavily supervise the bank accounts and transactions of users, albeit using automatic cloud services, removing the option to privately exchange money digitally. The transfer of money comes with disproportional costs, adverse cross-border payments in terms of speed and additional costs, and unwanted transparency. Many of these issues can be solved by the use of blockchain technology, and specifically Central Bank Digital Currencies (CBDC). With (almost) zero costs, transactions are executed between wallets anywhere in the world in a matter of seconds. Even internal use of blockchain for banks themselves will save about 10 billion dollars globally \cite{banks-blockchain2021}. Despite the openness of blockchain transactions to practically anyone, pseudo-anonymity is maintained as the identity is not revealed.

Centralized platforms inherently lack the self-sovereignty of data of their users. As the use of central servers is profitable in terms of availability and synchronization, personal data and metadata of users are stored. Although the data itself is often encrypted, the metadata contains valuable information (sender, recipient, time, location). WhatsApp is the most used messaging app \cite{messagingstats-2021} and promises its users end-to-end encryption. That does not withhold them to store a vast amount of metadata. Even though these platforms make us think that they prioritise security and privacy, not giving up their centralized nature is the primary reason for the existence of cyber attacks \cite{whatsappmanipulation-blackhat2019}. 

Decentralization as part of a societal infrastructure with a self-sovereign identity introduces many new opportunities and advantages. The user obtains a more centric position, has more privacy, and remains in control and in possession of their own data and identity. A self-sovereign identity can even serve a wider range of applications such as building authentic trust with others in online conversations. Communication no longer includes the storage of personal metadata on servers. Governments require less highly-secured infrastructures, which reduces the overall costs for both governments and citizens. The transfer of digital money between identities has major advantages. A reformed financial system is more efficient, faster, and optimal for cross-border payments while preserving the privacy of citizens.
In the following sections, we discuss the related work in \ref{Sec:Related-work}, the design and implementation of the first user- and identity-centric infrastructure is presented in sections \ref{Sec:Infrastructure}, \ref{Sec:Design}, and \ref{Sec:Implementation}. Our custom-designed P2P data transfer protocol is analyzed and evaluated in \Cref{Sec:ExperimentalAnalysis}.
\begin{table*}
\caption{Comparison with related works}
\begin{minipage}{\textwidth}
\scriptsize
\resizebox{\textwidth}{!}{
\begin{tabular}{c | l | c  c c c c | l c | c r}
\rot{\textbf{}} & \rot{\textbf{}} &  \rot{\textbf{P2P}}          & \rot{\textbf{open source}} & \rot{\textbf{E2E Encryption}}                          & \rot{\textbf{metadata}}     & \rot{\textbf{requirements}} &  \rot{\hspace{10pt}\textbf{\begin{tabular}[c]{@{}l@{}}attributes\\used for trust\\enforcement\end{tabular}}}                       & \rot{\textbf{wallet}}        &\rot{\textbf{maturity }\footnote{the current state of development in terms of completeness and usefulness}} & \rot{\textbf{note}} \\ \hline \hline
\multirow{11}{*}{\STAB{\rotatebox[origin=c]{90}{\small Centralized}}} & \textbf{WhatsApp} \cite{whatsapp-whitepaper2021}            &  {\color{DarkRed}\xmark} & {\color{DarkRed}\xmark}                                          & curve25519                                                     & {\color{DarkGreen}\cmark} & phone number           & \begin{tabular}[c]{@{}l@{}}phone number, name,\\ profile picture and status\end{tabular} & {\color{DarkRed}\xmark}          & high              &               \\ \cline{2-11}
& \textbf{Facebook Messenger} \cite{facebook-whitepaper2017}  & {\color{DarkRed}\xmark} & {\color{DarkRed}\xmark}                                          & curve25519                                                     & {\color{DarkGreen}\cmark} & Facebook profile       & \begin{tabular}[c]{@{}l@{}}Facebook profile,\\ name, profile picture\end{tabular}    & {\color{DarkRed}\xmark}                 & high              & \footnote{\label{footnote:e2e-default}E2E encryption not enabled by default}             \\ \cline{2-11}
& \textbf{WeChat (QQ)} \cite{wechat-2022}          & {\color{DarkRed}\xmark} & {\color{DarkRed}\xmark}                                          & {\color{DarkRed}\xmark}                                          & {\color{DarkGreen}\cmark} & phone number                           & \begin{tabular}[c]{@{}l@{}}phone number, name,\\ profile picture and ID\end{tabular}   & money               & high              &               \\ \cline{2-11}
& \textbf{Telegram} \cite{telegram-2022}            & {\color{DarkRed}\xmark} & {\color{DarkRed}\xmark}                                          & MTProto                                           & {\color{DarkGreen}\cmark} & phone number           & \begin{tabular}[c]{@{}l@{}}phone number, name, username,\\profile picture and status\end{tabular} & {\color{DarkRed}\xmark} & high              & \textsuperscript{\ref{footnote:e2e-default}}              \\ \cline{2-11}
& \textbf{iMessage} \cite{imessage-2021}            & {\color{DarkRed}\xmark} & {\color{DarkRed}\xmark}                                          & \begin{tabular}[c]{@{}c@{}}NIST P-256 curve\end{tabular}                                           & {\color{DarkGreen}\cmark} & Apple profile           & \begin{tabular}[c]{@{}l@{}}phone number, name,\\email, profile picture\end{tabular} & {\color{DarkRed}\xmark} & high              &               \\ \cline{2-11}
& \textbf{Signal Messenger} \cite{signal-messenger2017}    & {\color{DarkRed}\xmark} & {\color{DarkGreen}\cmark}                                          & \begin{tabular}[c]{@{}l@{}}curve25519,\\ curve448\end{tabular} & minimum \footnote{\label{footnote:metadata}no storage of metadata, only required for routing} & phone number                          & \begin{tabular}[c]{@{}l@{}}phone number, name,\\ profile picture and status\end{tabular}   & crypto        & high              &               \\ \hline \hline
\multirow{9}{*}{\STAB{\rotatebox[origin=c]{90}{\small Decentralized}}} & \textbf{Session Messenger} \cite{session-messenger2020}   &  {\color{DarkRed}\xmark} & {\color{DarkGreen}\cmark}                                          & \begin{tabular}[c]{@{}l@{}}curve25519,\\ curve448\end{tabular} & minimum \textsuperscript{\ref{footnote:metadata}}         & {\color{DarkRed}\xmark} &  \begin{tabular}[c]{@{}l@{}}name,\\profile picture\end{tabular}   & {\color{DarkRed}\xmark}                                                                              & medium            & \footnote{fork of Signal Messenger, onion routing for metadata anonymity, undelivered messages stored one of the distributed service nodes}             \\ \cline{2-11}
& \textbf{Status.im} \cite{status-whiterpaper2017}             & {\color{DarkGreen}\cmark} & {\color{DarkGreen}\cmark}                                          & curve25519                                                     & minimum \textsuperscript{\ref{footnote:metadata}} & {\color{DarkRed}\xmark}                 & \begin{tabular}[c]{@{}l@{}}username,\\profile picture\end{tabular}       & crypto                                                                          & high               &             \\ \cline{2-11}
& \textbf{Sylo} \cite{sylo-2020}      & {\color{DarkGreen}\cmark} & {\color{DarkRed}\xmark}                                          & curve25519                                                     & {\color{DarkGreen}\cmark} & {\color{DarkRed}\xmark}                 & \begin{tabular}[c]{@{}l@{}}name,\\profile picture\end{tabular}     & crypto & high              & \footnote{everyone can set up node and will be rewarded in crypto token SYLO}             \\ \cline{2-11}
& \textbf{Berty} \cite{berty-2020}              & {\color{DarkGreen}\cmark} & {\color{DarkGreen}\cmark}                                          & curve25519                                                     & minimum \textsuperscript{\ref{footnote:metadata}}               & {\color{DarkRed}\xmark}  & \begin{tabular}[c]{@{}l@{}}name,\\profile picture\end{tabular}  & {\color{DarkRed}\xmark}                                                                               & medium            &               \\ \cline{2-11}
& \textbf{Our design} (\Cref{Sec:Design}) & {\color{DarkGreen}\cmark} & {\color{DarkGreen}\cmark}                                          & curve25519                                                     & minimum \textsuperscript{\ref{footnote:metadata}}              & official Identity                     & \begin{tabular}[c]{@{}l@{}}identity name and verification\\status, profile picture\end{tabular} & crypto  & medium            &           \\ \hline   
\end{tabular}
}
\end{minipage}
\label{Table:platforms_competitors}
\end{table*}

\section{Related Work} \label{Sec:Related-work}
Self-sovereign identity provides citizens the control of their own identity. Citizens currently authenticate their identity to organizations and institutions, stored on the governments central server. DigiD\footnote{\url{https://www.digid.nl}} is the primary identity authenticator in the Netherlands and enables citizens to authenticate. Personal data is transferred from the government's server to the organization's server. This requires a perfectly secure connection and infrastructure on both sides. The citizen has no control over what information is shared. Additionally, these authentication mechanisms are disproportionately expensive and credited \euro$0,13$ for all 545 million successful authentications in 2021 \cite{digidcosts-2021,digidusage-2022}, especially due to increase of COVID-related authentications. Self-sovereign identities can be applied to mobile applications that replace authentication services like DigiD at a fraction of the cost. The first example is IRMA\footnote{\url{https://irma.app}}, an operational mobile platform that fetches the identity from the governmental servers once and stores the SSI and other personal information locally on the phone. The authentication to organizations can instead be performed offline using the stored SSI. The user remains in control by seeing the requested and shared information. Sovrin Network\footnote{\url{https://sovrin.org}} uses the same methodologies but instead enables other developers to build their own SSI application on top of their blockchain-based ecosystem. Some situations require the option to revoke the self-sovereign identity, for example when the identity document is lost or stolen. Both IRMA and Sovrin apply (centralized) authorities to revoke identities. This is a violation of the principles of SSI as it should be an authority-free system. The work of \citet{ssi-rowdy2021} incorporates distributed attestations for self-sovereign identities that includes offline revocation. Offline verification of attestations offers increased privacy and robustness. 

As mentioned before, this paper presents a \textit{novel} decentralized infrastructure and incorporates a government-issued identity within a messaging platform. Many other platforms exist, both centralized and decentralized, that apply at least some of the key points of this paper. \Cref{Table:platforms_competitors} portrays a (non-exhaustive) list of significant competitors in the market. The difference between the centralized and decentralized platforms shows a clear clustering. All centralized platforms require a privacy-sensitive asset and attributes that are shared with contacts to identity and enforce trust. The decentralized platforms are examples of Privacy by Design \cite{privacybydesign-Cavoukian2010} as they try to minimize the (centralized) storage of data and leakage of privacy-sensitive information. Decentralized platforms have not explicit initial requirements and the trust attributes are limited to manually chosen names and profile pictures. 

\begin{figure*}[ht!]
    \centering
    \includegraphics[width=0.7\textwidth]{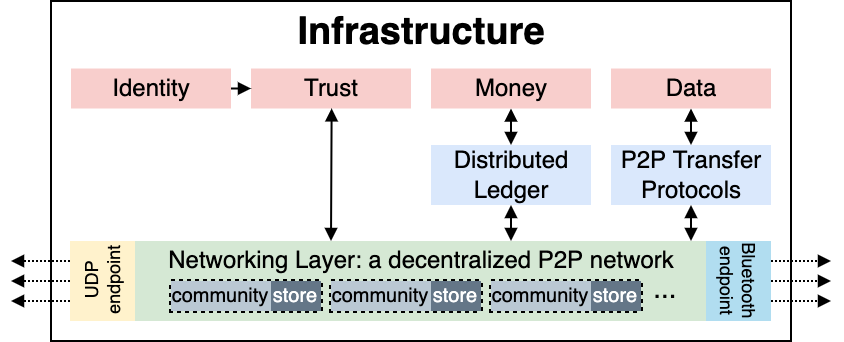}
    \caption{General overview of the infrastructure}
    \label{Figure:infrastructure}
\end{figure*}

WhatsApp \cite{whatsapp-whitepaper2021}, Facebook Messenger \cite{facebook-whitepaper2017}, and WeChat \cite{wechat-2022}, are fully centralized platforms that store metadata of their users. The industry-standard E2E encryption curves are used by all platforms but Telegram \cite{telegram-2022}, that uses their self-designed protocol, and the unencrypted WeChat. WeChat, which is monitored by the Chinese government, incorporates strong censorship and interception protocols for data exchanged by its citizens. This degree of monitoring is not present in other (centralized) platforms. Governments even oblige these centralized platforms to share their stored metadata or apply censorship in some situations \cite{metadata-2021,rollingstone-2021}. Signal Messenger \cite{signal-messenger2017}, which is centralized but specifically designed with privacy in mind, does not store any personal information. However, central servers are necessary for routing and functionalities like account recovery. Account activation is validated using the phone number or email address of the user. Characteristically, most of the centralized platforms don't provide full transparency and do not disclose the complete structure of their platform. 

Decentralized infrastructures try to realize anonymity by minimizing the metadata in the network. Session Messenger \cite{session-messenger2020}, a \textit{decentralized} fork of Signal Messenger, attempts to provide anonymity and preservation of privacy using onion routing \cite{onion-routing1998}. It makes it nearly impossible for any intermediary (node) to derive both the sender and receiver of the message. Onion routing is not suitable in a (fully) P2P network as peers only know a limited number of other peers and do not (necessarily) communicate with nodes. Status \cite{status-whiterpaper2017}, Sylo \cite{sylo-2020}, and Berty \cite{berty-2020} are decentralized, P2P, secure, minimize leakage of privacy-sensitive information, and have no initial requirements. Status is build on the Ethereum network and incorporates its own utility network token to provide paid features to users. Like Session Messenger, undelivered messages are stored on nodes that obtain your IP address to deliver it later. This characteristic contradicts the principles of privacy. Sylo is a fully operational platform that is not protected against possible leakage of metadata and does not provide full transparency to its users. Berty is secure and transparent, minimizes leakage of privacy-sensitive information in terms of metadata and requirements, and therefore suits our requirements best, but is currently still underdeveloped. 

There are many implementations with the same design characteristics. The idiomatic platform is decentralized, does not require temporary storage of messages on nodes due to its P2P nature, incorporates trusted encryption, does not require and store metadata, and has no redundant identifiable requirements. Our design uniquely adds a self-sovereign identity, providing various functionalities. To achieve higher trustworthiness, trust enforcement attributes must not only consists of manually forge-able components. Furthermore, the platform must contain private and secure mechanisms for the transfer and storage of data and digital money.
\section{Infrastructure} \label{Sec:Infrastructure}
The dominant problem of market-leading societal platforms is their centralized nature. Decentralization targets many weak spots of centralization, such as providing private and secure storage of data in a distributed way. Even a decentralized network allows the storage of metadata by the nodes that are traversed on its path to the destination. Some nodes may even sell metadata to third parties. A P2P network enables direct communication with peers without any intermediary. As no intermediary is able to act as a middle-man or adversary, it serves as an extra layer of protection against malicious or intentional behavior. The communication is only secure if the message, or data, is encrypted. 

We also require a component to store and exchange data in a distributed manner. One of the requirements of distributed systems is synchronization across many independent nodes. This is difficult to realize in systems that include peers that are not connected at all times. To enable the transfer of (digital) money, the infrastructure requires a persistent and decentralized storage of data that does not require continuous synchronization for all peers. Blockchains store transactions between two wallets in a permanent and uneditable manner. Every transaction on the blockchain is entangled to its previous block, making it a reliable chain of tamper-proof assets. This form of storage is a fast, lightweight, and structured alternative to conventional storage, albeit visible to everyone. Every transaction can be back-traced to create a well-organized overview, which is well suited to serve as a wallet. The infrastructure additionally requires data transfer protocols to exchange data. \Cref{Figure:infrastructure} portrays a low-level overview of our infrastructure.

Peers constantly observe other peers in the network. The infrastructure requires anonymous peer identification as it is not desirable to spread personal information to (unknown) peers in the network. To ensure a secure communication channel, the principles of the CIA Triad \cite{ciatriad} must be in place. The objectives of a secure system include \textit{Confidentiality}, \textit{Integrity}, and \textit{Availability}. Confidentiality ensures that information is only accessible to authorized parties. Digital signatures ensure the integrity of the information by providing proof that it originates from the sender and has not been altered by any third party. Availability ensures that information is available to authorized parties at any point in time. In P2P networks, only the last principle is difficult to realize due to its dependence on the connectivity of individual peers.

Confidentiality requires the encryption of information. Public-key cryptography \cite{public-private-crypto} is the most commonly-used mechanism for secure communication. Confidential exchange of messages and data additionally enables the identification of peers exposing private information. Each peer has a public-private key pair. The private key is cryptographically generated once and only be known to the owner. The private key decrypts the encrypted data. The public key is mathematically derived from the private key and may be public as it is computationally infeasible to derive the private key from the public key. The public key provides several applications in our infrastructure. Foremost, the public key of the intended recipient is used for the encryption of the data. Secondly, digital signatures prove the authenticity of data and can be verified using the public key of the signatory/sender. And lastly, peers can be identified and distinguished using their public key. 

\subsection{Networking Layer}
Our infrastructures requires a networking layer to handle outgoing and incoming communication with peers. A P2P networking layer that is authenticated and privacy-aware is IPv8 \cite{ipv8-2020}. IPv8 is developed as an academic successor of IPv4 and attempts to overcome IPv4's weak characteristics and growing number of problems. The objective of IPv8 is to provide communication in a zero-server infrastructure, equal status and power within the network for everyone, and perfect secrecy with E2E encryption. IPv8 can establish connections to peers, even for devices connected behind NAT or a (strong) firewall. The endpoints of the networking layer are independent of any central infrastructure.

The aim is to minimize the exposed metadata. For this reason, data packets can't be broadcasted on the network, in the hope that other peers can deliver the packet to the intended recipient. Some P2P systems employ distributed nodes to deliver the data in this situation. These nodes temporarily store data and metadata until delivered. Although this addresses the availability of information in the network, it exposes the storage and possible leakage of metadata. The metadata of a packet should (ideally) only contain information about delivery, that is, the receiver's public key or IP address. The metadata that is not relevant for delivery should be encrypted with the data. The risk of exposing privacy-sensitive information in our P2P network is minimized as peers directly communicate without serviceable nodes or peers. Peers change connectivity status or change their network address regularly. In these situations the peers announce their new address to all former peers to keep the previous connections alive. Peers also connect to random peers to broaden their network reach. While this may sound contradictory, it does not violate their privacy. No personal information, including the intentions and previous communication histories, can be deduced.

IPv8 applies the concept of network overlay or \textit{community}. This enables developers to build applications on top of the base networking layer by creating their own community. We need several communities in our infrastructure to satisfy our requirements. The base community includes all functionalities related to peer connectivity, communication, data serialization, and encryption. The IPv8 networking layer combines responsibilities from several layers of the OSI model \cite{osi-model1994}. Communities that require the storage of data, can use an additional \textit{store} that handles the interaction with a database. A discovery community handles the discovery and connection of (new) peers present in the same community. Every community and peer may have a different list of connected peers. The communication between peers in the network and community is handled by endpoints/sockets. IPv8 provides support for both online and offline communication using UDP and Bluetooth endpoints. The support for offline communication increases the reliability and applicability of the platform, especially in areas with low network coverage.

\subsection{Distributed Ledger}
\begin{figure}[ht!]
	\centering
	\includegraphics[width=\linewidth]{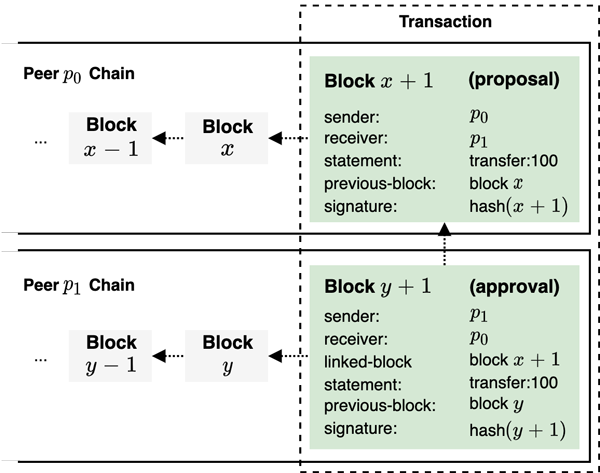}
	\caption{Transaction in the ledger}
	\label{fig:transaction}
\end{figure}

The infrastructure requires a distributed ledger that handles and stores the financial transactions. The permission-less and scalable distributed ledger TrustChain \cite{trustchain-2020} is already integrated as a community in IPv8. TrustChain is capable of sending and receiving trusted transactions between peers. The blockchain-based data structure is a tamper-proof immutable chain of transactions. No central authority has control over the transactions. Every peer implements a personalized chain that contains only blocks related to that peer, i.e. either sent or received by the peer. TrustChain has three basic functionalities: transmit/receive blocks, broadcast blocks, and chain crawling. The transmit and receive process merges both parties in one transaction, see \Cref{fig:transaction}. 
The initiator $(p_0)$ creates, signs, and sends the \textit{proposal block} to the counterparty $(p_1)$. On receipt of the \textit{proposal block}, $(p_1)$ creates, signs, and sends the \textit{agreement block} back to $p_0$. Both blocks are linked by the public key of the counterparty and can be considered as half blocks that together form a transaction. During the process, the integrity of the received blocks is validated and both parties add the half blocks to their chains. The transaction is complete when both parties received and signed both half blocks, i.e. both acknowledged and accepted its contents. The broadcast functionality enables to transmit the block to all connected peers. Chain crawling is the retrieval of a peer's chain using its public key. 
\section{Design} \label{Sec:Design}
We can divide the design of the platform into four pillars: identity, trust, money, and data. These elements are integrated within the infrastructure of \Cref{Sec:Infrastructure} to create an ecosystem that combines these elements seamlessly. The elements, described in the following sections, must satisfy the requirements and functionalities that are deemed necessary for a self-sovereign, secure, and privacy-aware communication platform.

\subsection{Identity} \label{Subsec:Design-identity}
Identity is an integral part of citizens when it comes to ownership over their self-sovereign identity. Integration of government-issued travel documents in a self-sovereign manner introduces various new opportunities. One of the major applications is the authentication to online organizations. The citizen controls the exchange of its own identity information to organizations instead of the conventional online governmental authentication that blindly transfers all available information. Authentication can only serve its purpose if the information within the self-sovereign identity is authentic. IRMA, the application mentioned in \Cref{Sec:Related-work}, achieves authenticity by fetching the identity information from the government's central server once. As we desperately want to eliminate external dependencies, this method is not suited to our system. Every (adult) citizen is obliged to legitimate himself with a passport or identity card upon request by the authorities. As it is mandatory to posses such a document we can apply these documents as an authentic base for our platform. The identity written in the machine-readable zone (MRZ) of the document, contains the same personal information as on the government's server.

\begin{figure}[ht!]
\centering
  \subfloat[a][\centering Scan MRZ zone of document\newline using the camera of the phone]{\includegraphics[width=0.5\linewidth]{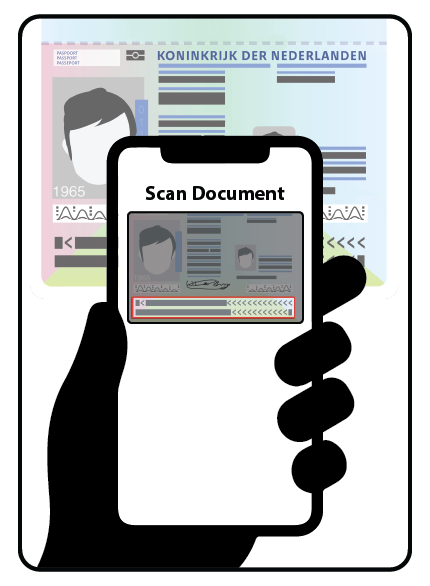} \label{subfig:document-scan}}
  \subfloat[b][\centering Reading biometric chip using the\newline NFC chip of the phone]{\includegraphics[width=0.5\linewidth]{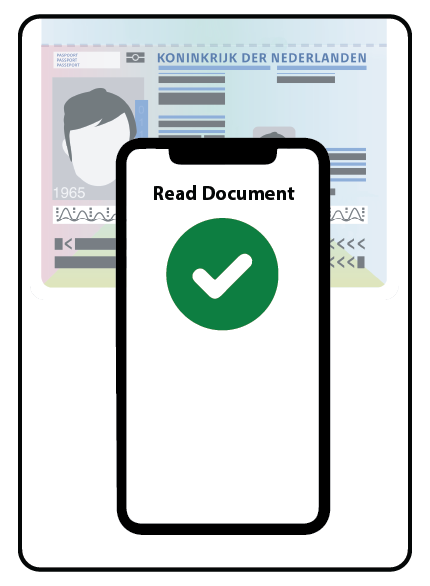} \label{subfig:document-read}}
  \caption{Identity document onboarding process}
  \label{fig:document_onboarding}
\end{figure}

The identity document onboarding process is executed in two consecutive steps as in \Cref{fig:document_onboarding}. In the first step the user must scan the text in the MRZ of the document using the camera of the device, as in \cref{subfig:document-scan}. A combination of AI and check digits (embedded in the MRZ) ensure a syntactically valid, not necessarily correct, recognition of the identity information that is saved on the device.
The second step is extremely important as it proves the authenticity of the document digitally. This validation step determines the correctness of the scanned attributes. Most current documents are equipped with a built-in biometric chip in which the information is embedded. The NFC chip of devices enables them to communicate with the document. The device must steadily be placed with its back to the document until all information has been transferred, see \Cref{subfig:document-read}. The biometric chip contains multiple layers of protection against for example eavesdroppers and modification. The protection prevents the unauthorized reading of the document using for example NFC skimmers. The authentication requires the document number, date of birth, and date of expiry as passwords, as obtained in the initial step. After the connection has been established, the biometric chip transfers all requested attributes to the device. All stored attributes together form the self-sovereign identity of the user. The onboarding process is deemed authentic and secure because (I) a physical document is required and the attributes displayed on the card must match the content in the biometric chip of the card, and (II) the biometric document is widely applied for governmental purposes and international traveling, without excessive vulnerabilities \cite{biometric-passport-attacks2020}. One issue that remains is the possibility to revoke access to a self-sovereign identity. Millions of documents get stolen or lost yearly \cite{identity-lost2022}, and risk of being used by others. There is no way to recognize and revoke access without knowledge from a central server. The self-sovereign identity must be valid, just like physical documents, until the expiration date of the document.

A different situation arises when the phone has no support for the NFC chip, defectively or physically. There does not exist an (offline) method to obtain the identity authentically by only scanning the document. As the biometric chip performs the validation of the document, a malicious actor can forge the complete MRZ to its preference. No identity-related functionalities can be trusted to contain truthful and authentic information. There is no other option to either disable all these identity-related functionalities for these devices or to be dependent on the government's central server to obtain the identity.

In real-world situations, it is sometimes mandatory to show or even make a copy of your physical identity document to verify your identity or to serve as insurance. The authority is not only capable of unnecessarily viewing the requested attribute(s), but also other attributes on the document. This is a direct violation of citizens' privacy and can even lead to identity theft or misuse of a person's integrity. Citizens are forced to trust this authority to handle and store their identity secure and with care. Self-sovereign identities introduce the opportunity to use verifiable claims. Verifiable claims are claims about information that can be verified using attestations. \citet{ssi-rowdy2021} designed a framework that incorporates revocable verifiable claims without revealing the actual requested piece of information using zero-knowledge proofs. To apply variable claims in a trustworthy manner, the information from the self-sovereign identity must, again, be authentic.

The focus should not only be on data in transit, but also on data at rest. In the latter, the data is stored somewhere without anyone currently accessing it. The data is often stored as a file or in a database. In our case, the identity must be secured significantly without risking identity theft. As mentioned before, our design applies public-private keys for encryption. The storage of the identity can easily be encrypted using the private key of the user, while only allowing the application to decrypt when absolutely required. That means that no one is able to access the contents outside the application environment as long as the private key is insusceptible. Biometric protection (face recognition or fingerprint) or the use of passcodes, can be applied as another layer of protection against unauthorized access to sensitive data. It can also serve as confirmation for irreversible actions like the transfer of sensitive information or money. 
\subsection{Trust} \label{Subsec:Design-trust}
Platforms have to deal with several types of trust. The first natural form is trust in a system or platform. This is the case for nontransparent centralized platforms. As a user, you want to have faith that your personal data is handled and stored with care. This is often one of the primary problems with centralization. As all user data is stored on the platform's servers, you must have confidence that the data, including metadata, is protected with the highest security standards, exchanged in encrypted form, and not sold to any third parties. If no good alternative platform exists, or because no friends use other platforms, the user has to decide whether to continue to use the platform and neglect the privacy-related issues, or not use the platform anymore. Often the first choice is selected as people value the use of the service more than their own privacy. Decentralization (almost) completely eliminates this trust, or distrust, as there is no central component or authority that decides over you and your data. Also, a platform that is open source is generally more trusted as there are no hidden surprises due to the reviews of experts. We must take notice that in some networks, malicious actors actively crawl metadata in order to gain knowledge about confidential information. Communication in P2P networks, in the form of network packets, is directly transmitted to the network address of the recipient, making this problem almost redundant.

In messaging and societal applications another form of trust arises: the trust in the real identity of the other participant. The confidence in the authenticity of the contact is determined based on various factors like the (online) identity of the contact, the (dis)similarity in the way they communicate, and the discussed topics. The style of writing and difference in for example punctuation and the use of capital letters may also play a role in recognition. Unfortunately in most applications, personal information can easily be forged or stolen from people's (real) online identities. If we again look at \Cref{Table:platforms_competitors}, most attributes for trust enforcement of centralized platforms are forgeable. Spear phishing \cite{spear-phishing2019} is an cyber attack in which individuals are targeted with the explicit use of personal information to gain knowledge or access to (more) sensitive information. In applications like WhatsApp, it is possible to migrate from one phone to another. As this is a convenient feature, it also exposes the risk that hackers can take over your account on their phone and communicate with your contacts instead. This has the deficit that hackers are able to use these accounts to steal confidential information or even request money from trusted unsuspecting contacts. These hackers attempt to mimic as much confidential information of the hacked person to not arouse suspicion or to simply gain trust with their new victims. For this reason we did not include migration in our platform (yet). Another situation arises when the phone is stolen or lost. As we don't have control anymore, it is (similarily to other platforms) possible to impersonate someone on a lost or hacked phone.

\begin{figure}[ht!]
	\centering
	\includegraphics[width=.9\linewidth]{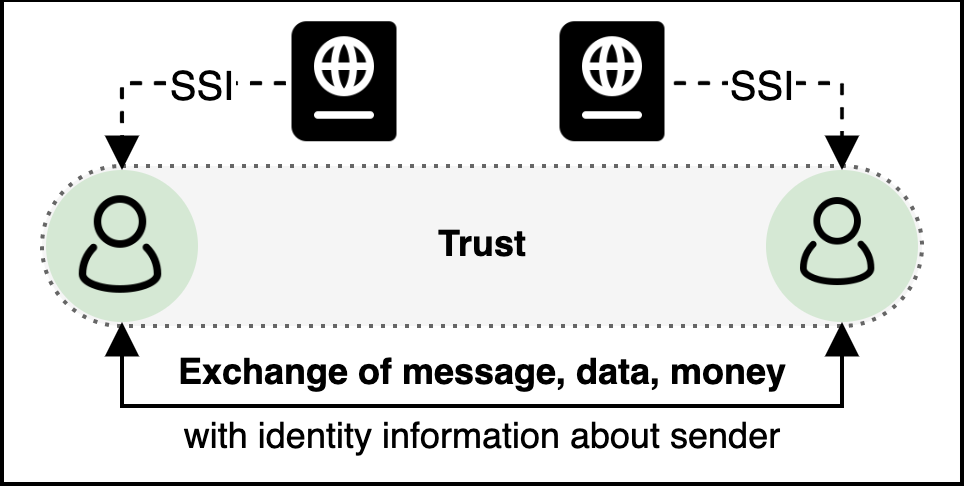}
	\caption{Building trust using the self-sovereign identity}
	\label{fig:ssi_trust_enforcement}
\end{figure}

The challenge is to exchange the right amount of trust to the recipient of your message without excessively exposing private information. In the initial phase of the conversation, especially if the users connected in some online way, trust (or distrust) can play a major role. As valid self-sovereign identities are incorporated in our design, we can access and apply this authentic information. In a normal, physical first meeting, one would introduce themselves by their (first) name, and indirectly with facial expressions, the sound of their voice, and the overall atmosphere. These aspects are not available in the digital world without extra efforts. The only identifiable information that we can exchange is a person's name and photo as embedded in the self-sovereign identity. These attributes can provide the authenticity that our system requires by sending it to the other contact. To provide proof of its authenticity, we can include the identity verification status. The verification status formally denotes the trustworthiness of the name and photo, while it technically denotes the use of the biometric and NFC chip. A different situation arises when users can't use the NFC chip. These users can't be verified as a result, and are able to choose their own name and photo to allow them access and use of the platform. The contacts will in turn be notified of the unverified status, implicating they should act cautiously. 

\begin{table}[ht!]
    \centering
    \caption{Trust enforcement combinations for identity name}
    \label{Table:Trust-name-options}
    \begin{tabular}{l l l}
         \multicolumn{2}{l}{\textbf{Combination}} &  \textbf{Example} \\\hline
         I & \{First Name\} & Timothy John \\
         II & \{Last Name\} & Berners-Lee \\
         III & \{First Name\} \{Surname[0]\} & Timothy John B.L. \\
         IV & \{First Name[0]\} \{Surname\} & T.J. Berners-Lee \\
         V & \{First Name\} \{Surname\} & Timothy John Berners-Lee \\ \hline
    \end{tabular}
\end{table}

Of all information in the self-sovereign identity only the name and picture, and possibly the age and gender, are suitable to enforce trust. The transfer of too much confidential information can lead to malicious misuse of them or their contacts. We should therefore limit the exchange of information. We can compose various combinations of the given and last name that aim to provide trust, see \Cref{Table:Trust-name-options}. While combination \texttt{I} and \texttt{II} are too general and unidentifiable, the use of the full name in \texttt{V}, as identically embedded in the self-sovereign identity, is an easy source for malicious actors to take advantage of. The combinations \texttt{III} and \texttt{IV} include both the first and last name in a modified form and provide a more personal and identifiable view without exaggerating. The identity of a person is more decisive by its last name due to its uniqueness, and therefore combination \texttt{IV} fits our purpose best for the use as the trust enforcement attribute along with the picture. The exchange of the age and gender has been considered, but increases the risk of impersonating.

The name and photo attributes of the peer's identity are sent encrypted along with every message. Upon receipt, the system is able to detect differences with the information of the currently stored state. Initially, the state is empty. The recipient of the first message will be notified in a recognizable manner that the identity of the contact has been determined. For every other message, if at some point the state changes, the user will receive similar notifications stating that the information has been updated. This mechanism makes sure the user always has knowledge of the sender's \textit{formal} identity. It is, however, impossible to notice difference in case a phone is stolen or accessed unauthorizedly without an alteration of the identity. On-device authorization is desirable using for example biometric protection. As long as biometric protection is in place, it should be difficult to impersonate. Also, a mechanism that requires the user to regularly verify its identity using their physical document could help to reduce misuse. Both options are currently not part of our initial design but serve as improvements on authentic communication. It is equally important to not only focus on building trust but to preserve the privacy of the receiver as well. The platform will \textit{never} share identity information without having sent a message or transaction first. This reduces the risk of malicious actors purposely attempting to fetch names and photos linked to particular public keys.
\subsection{Money} \label{Subsec:Design-money}
As the need for financial privacy grows, many Web3 applications integrate the transfer of some sort of value in the form of cryptocurrencies or NFT's. Governments of nations and unions are currently exploring the economic and technical feasibility of Central Bank Digital Currencies (CBDC) \cite{cbdc-fed2022,cbdc-ecb2021,cbdc-pound2020,cbdc-yuan2021}. Money in the form of digital currencies that reflect on their native currency, often also referred to as stable coins, could provide a fast, cheap, and private exchange between participants. Other cryptocurrencies are not suited for this purpose as they appear to be extremely volatile, therefore lacking the consistency for a safe store of value. CBDC has three main characteristics: it is a digital currency, it is issued by a central bank, and it must be universally accessible. As these currencies will be legally recognized and backed by their governments and central banks, their introduction and effect on the financial system are heavily tested. If somewhere in the future, countries decide to become cashless, these currencies must be creditable to replace coins and banknotes. China is already in the advanced stages of the development of its CBDC and is testing the implementation of its digital Yuan wallet in a second pilot \cite{cbdc-yuan-pilot2022}. The Bank of China will still include regulation for larger transactions and seek only to collect personal information that is legally required. Compared to the principle of cryptocurrencies, which strives for pseudo-anonymous transactions, China doesn't make an effort to give up its (financial) regulation. 

Governments, banks, and tax offices shouldn't have implicit insights into transactions of CBDC's. The privacy of the users will be preserved up to a certain level. As most ledgers and blockchains are transparent, transactions on the chain are visible to others, and can even view or attempt to trace back the wallet balance. Blockchain still provides pseudo-anonymity because participants of transactions are often only identified by a public key. No further personal information is attached to transactions apart from the sender and recipient and some unidentifiable transaction contents, statistics, and possibly some other (encrypted) data. Governments will not attempt to regulate and gain insight into these transactions because, similarly to cash, it is simply not feasible to do so. If we compare blockchain transactions with current digital payment solutions, it is definitely a step forward, while conventional cash remains the most private and anonymous form of payment and store of value. Not only would the use of CBDC's contribute to new innovations and direct accessibility of money without any external dependence, it may even serve some of the fundamental financial primitives (lending, borrowing, liquidity).

Currently, many external services or banks provide the functionality to create payment requests. Its creator shares a link to all participants that redirects to the payment portal of the service. This not only creates an additional dependency on the use of a centralized (paid) service but also opens abusive opportunities for malicious actors. Many fraud cases \cite{tikkie-fraud2020} using payment requests make the service vulnerable, especially for unsuspecting persons or the elderly. P2P digital payments can solve this problem by eliminating the dependence on the middle-man. The transaction or payment request is instead directly sent to the other peer, in an online or offline fashion. This not only makes it faster and cheaper but also offers a more private exchange of value.

Our design incorporates an existing implementation of an offline-capable euro CBDC called EuroToken \cite{eurotoken-blokzijl2021}. It utilizes the distributed ledger TrustChain \cite{trustchain-2020} that builds upon the technologies of the IPv8 network \cite{ipv8-2020}. The EuroToken protocol tries to offer a scalable, privacy-aware, and cheating-resistant system for the exchange and storage of transactions. Every block stored on the ledger contains a single transaction that states the transfer of funds from one to another. A block is cryptographically linked to its predecessor and therefore preserving a chain of chronological and valid blocks. Transactions are generally settled within seconds, independent of within- or cross-border payments, but require the connectivity of both parties to completely settle the transaction. 

One of the aspects of a tokenized system is the acquisition of tokens. For the system to be useful, it requires at least one option to buy and sell these tokens. EuroToken incorporates a \textit{central} exchange portal that allows users to exchange money on their bank account with EuroTokens, in both directions. The user and portal included in the exchange have to announce themselves to each other using their public key and additionally the transfer amount. The portal creates a request to the user that is automatically handled and stored by the protocol. Although it is not desirable to integrate centralized components in the system, it is considered an exception for the system to be functional. Improvements on the protocol could include distributed portals that are managed by random peers instead of a single entity.

After the initialization of the application, the wallet is ready to send and receive tokens. A balance is obviously required to send tokens. Tokens are either received by other peers or acquired using the exchange portal. The balance of a wallet is determined and validated using the blocks on the ledger. Our design fully integrates and stimulates the transfer of money between peers as it can be transferred directly from the wallet or indirectly from within a conversation with another peer. The integration of internal payment requests conveniently enables peers to request tokens from other peers. A transfer request differs from a transaction as is not formal and binding, and only contains the amount and the public key of the requestor.
\subsection{Data} \label{Subsec:Design-data}
One of the key applications of secure and private communication is the decentralized transfer of data. Data is the collective name for everything that can be expressed in the form of human-unreadable blobs, a Binary Large Object. These blobs can in turn be deserialized into a format that is readable for humans, e.g. images or text documents. Messages, and even transactions, can thus be classified as data as well. The current implementation of IPv8 contains a basic data transfer protocol. This protocol is able to send blobs to other peers, containing metadata and data. The transfer of data in form of messages and small blobs is fast but unreliable. However, the protocol has proven to be limited in terms of performance for larger-sized blobs. For proper use of our platform, it was deemed necessary to design a custom data transfer protocol that provides performance and reliable exchange of data, in a secure and private fashion.

\begin{figure}[ht!]
	\centering
	\includegraphics[width=.98\linewidth]{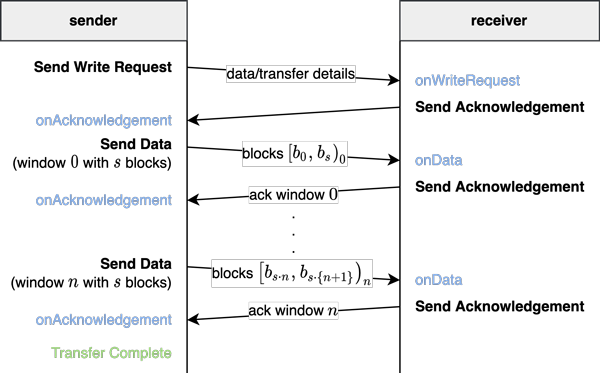}
	\caption{Normal operation of the data transfer protocol}
	\label{fig:eva_simple}
\end{figure}

The designed data transfer protocol aims to provide reliable and optimal performance for everybody in a \textit{progressive} and \textit{adaptive} fashion. The protocol is fully integrated into IPv8 and available to every community if necessary. Due to limitations, IPv8 (currently) only allows one concurrent transfer between two peers. The protocol is based on the principles of TFTP \cite{tftp-1981}, the Trivial File Transfer Protocol. TFTP is a simple and connection-less data transfer protocol, with the consequence that it is unauthenticated and no security mechanisms are provided. The transfer of confidential data in (external) networks is unsafe, and therefore not recommended. Dedicated connections with other peers cannot be established in P2P networks due to the lack of end-to-end connectivity. The long-existent TCP protocol is connection-oriented and much more reliable, but is not suitable for that reason. For a connection-less protocol, the application of the User Datagram Protocol (UDP) \cite{udp-1980} is an obvious choice. The question is how to effectively integrate the unreliable UDP in the design of a reliable data transfer protocol. For some purposes, for instance live video streaming, the loss of single packets does not impact the result as single video frames or pixels are simply skipped. In the case of our platform, the loss of packets would have the deficit of distorted and unusable data. Our protocol must therefore keep track of unreceived packets and request retransmission. The operation of the protocol is very basic and only consists of four different packet types. All packets are encrypted and in principle only decryptable to the intended receiver. The normal operation of the protocol is portrayed in \Cref{fig:eva_simple}. 
The sender of the transfer first has to request to write data by sending a \texttt{\textbf{WriteRequest}} payload/packet to the receiver. With this packet, the sender additionally announces all transfer- and data-specific details to make sure both parties have the same understanding. The receiver confirms this request by returning an \texttt{\textbf{Acknowledgement}} packet. This acknowledgment triggers the sender to start the transmission of data with \texttt{\textbf{Data}} packets. The data cannot be sent in one piece for multiple reasons. Firstly, the maximum UDP packet size is strictly limited to 1500 bytes due to the Maximum Transmission Unit (MTU) of the Ethernet \cite{mtu-tanenbaum2011}. The IPv8 protocol additionally requires a header of approximately 177 bytes to each block for routing, identification, and security purposes. Our \texttt{\textbf{Data}} payload header also requires identifiable information in the form of a block number, nonce, and some other attributes. This means that the data inside the packet can be roughly somewhere between 1200 and 1250 bytes. Secondly, since UDP is unreliable and packets are not guaranteed to be delivered, the transmission of the data at once (if technically feasible) would be too much of a gamble to arrive, especially for large blobs. The protocol is required to split the data into small(er) pieces to fit the packets, creating \textit{blocks}. Each of the blocks has a particular size in bytes and all blocks concatenated in the correct order represents the data. To reliably transfer data from one to another, we have to confirm the receipt of the data packets by again sending an \texttt{\textbf{Acknowledgement}} packet. To not unnecessarily wait for confirmation and delay the transfer, the acknowledgment (and any other packet) must be received within a certain interval before the previously sent packet is retransmitted. The principle of windowing allows multiple packets to be sent at once without requiring an acknowledgment for every single packet. This increases the performance majorly as most of the idle time of the sender is spent waiting for confirmation. The window size of the transfer defines the hard limit on how many bytes or, equivalently, the number of blocks if we take the block size out of the equation, can be sent within every window without intermediate acknowledging. After the last block of the window has been received, the protocol sends an acknowledgment to the sender. It does not necessarily mean that all blocks within that window have been received, as some arrive later and some will not be delivered at all. In that acknowledgment, the receiver includes the block numbers that have not been received (yet). The protocol could decide to only transmit individual packets and wait for the confirmation of receipt. This, however, has several disadvantages. Firstly, the transfer speed is significantly decreased as additional transmit and acknowledge stages are added, including waiting time. Regularly UDP packets arrive late or not at all, meaning that for a good part of the windows the unreceived blocks have to be retransmitted, even for single unreceived blocks. Secondly, by staying at the current window, it may occur that some of these packets have trouble being delivered, and the transfer may not progress further for a longer period (in terms of transfer time). To account for these drawbacks, the protocol will always try to \textit{progressively continue} its normal operation. This means that these unreceived blocks will piggyback with the next transfer window in the hope to be delivered without causing an overall delay. For every next window that passes, the confidence in these blocks being delivered increases. If all windows are sent, it may occur that there are some unreceived blocks left. In this case, the protocol will remain in the transmit phase of the last window until all blocks have been confirmed. The transfer is considered complete, for both the sender and receiver, after receipt of the acknowledgment (sender) and last block (receiver). 

The receiver may not be able to adhere to the write request because either the data does not comply with the allowed size between zero and a predefined limit, or both peers try to start a transfer at the exact same time. In these situations an \texttt{\textbf{Error}} packet is returned that contains the reason of refusal. Both the sender and receiver have no other option but to terminate the transfer. During the transfer, it regularly happens that no response is received for sent packets. Both the sender and receiver therefore use their own retransmit interval and retransmit attempt count. A retransmit is scheduled if the time between the last transmit and the allowed interval is exceeded. We don't want the protocol to retransmit infinitely during disconnectivity. This is prevented using the attempt count that only allows a maximum number of \textit{consecutive} retransmits. If after retransmission the other peer suddenly responds, the count is reset and the protocol continues normal operation. If the peer appears to be unresponsive, and the number of consecutive retransmits exceeds the attempt count, the transfer is considered timed out and will be terminated. In worst-case scenarios, often where the connection is unreliable or slow, the transfer of packets and confirmations may take too much time and can timeout the transfer. These connections could probably benefit from a lower window size as fewer blocks have to be transmitted and confirmed within the same time. The protocol \textit{adaptively} downscales the window size of a transfer after every timeout to give slower connections a better chance of success. Although it is important that the protocol is suitable for everyone in every situation, it is undesirable to lower the performance for everyone, due to a minor number of failures. We, therefore, chose to initially apply the optimal transfer settings for every transfer and lower the performance when required. The default parameters of the protocol that provide an optimal performance are analyzed in \Cref{Sec:ExperimentalAnalysis}. An optimization to the protocol can include the application of sliding windows for a more dynamic and optimal performance.
\section{Implementation} \label{Sec:Implementation}
In Sections \ref{Sec:Infrastructure} and \ref{Sec:Design} we've discussed the infrastructure and design of the main components that form the basis of our novel platform. The use of both IPv8 and TrustChain has proven to be a valuable fit for our infrastructure. The already existing implementations are applicable to a certain level as they have primarily been developed to serve as separate proofs-of-concept. These implementations additionally lack refinement and general cooperability and applicability with the other functionalities. In this section, the contributions to our complete infrastructure are discussed, as well as the changes to the existing implementations, and how they are integrated into our platform to provide a well-designed and well-functioning platform. The implementation is additionally concerned with the UX and UI design of the platform. The platform has been integrated within the TrustChain superapp\footnote{\url{https://github.com/Tribler/trustchain-superapp}}, an Android mobile application developed in Kotlin that contains many different small applications built on IPv8 and TrustChain. The data transfer protocol is integrated into the IPv8 stack\footnote{\url{https://github.com/Tribler/kotlin-ipv8}} and is available to every community. 

\begin{figure*}[ht!]
\centering
  \subfloat[a][\centering Old exchange implementation]{\includegraphics[width=0.25\linewidth]{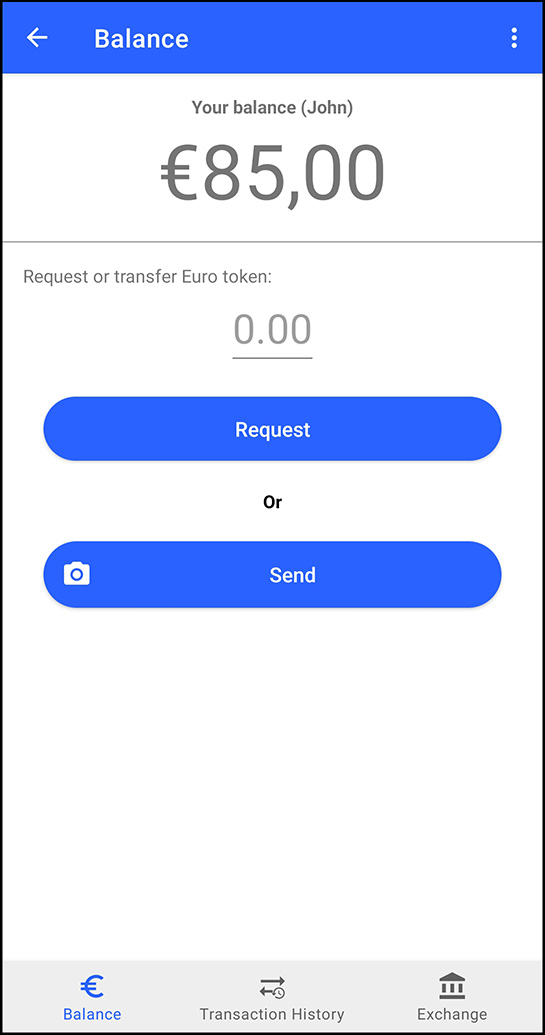} \label{subfig:implementatin_exchange_old}}
  \subfloat[b][\centering Our wallet overview (initial view) containing exchange widget \cite{timbernerslee}]{\includegraphics[width=0.25\linewidth]{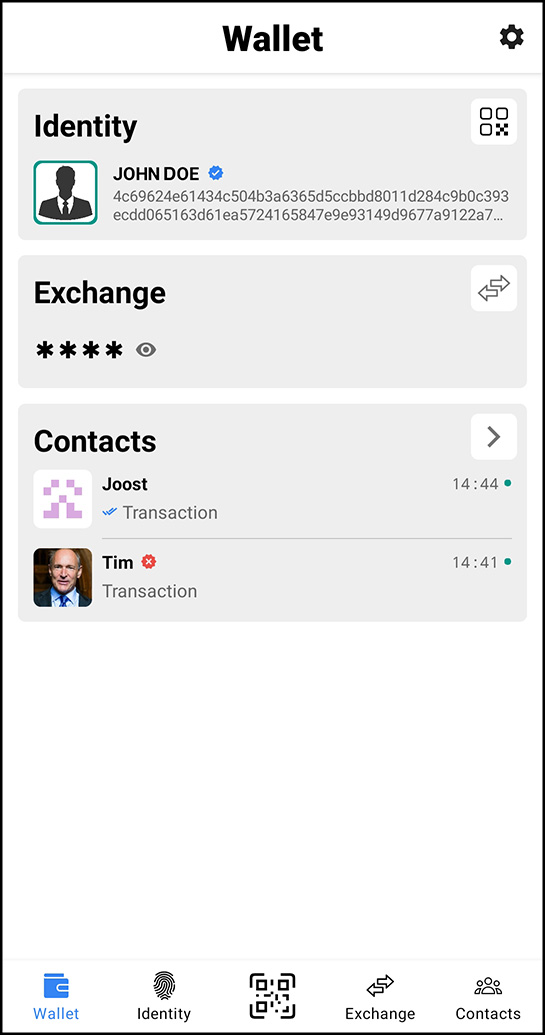} \label{subfig:implementatin_wallet_overview}}
  \subfloat[c][\centering Our exchange implementation and transfer options]{\includegraphics[width=0.25\linewidth]{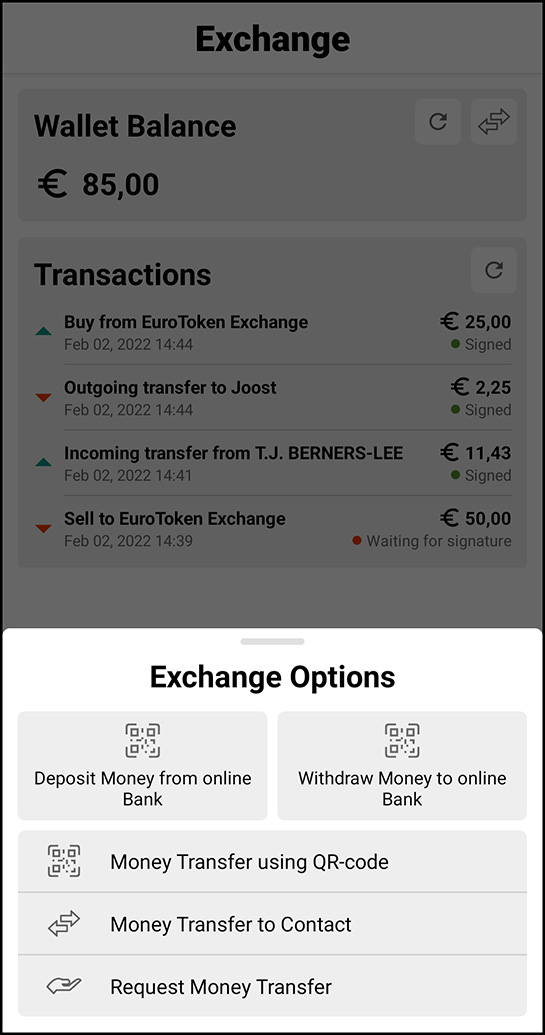} \label{subfig:implementatin_exchange_new}}
  \subfloat[d][\centering Our transaction detail view transaction \cite{timbernerslee}]{\includegraphics[width=0.25\linewidth]{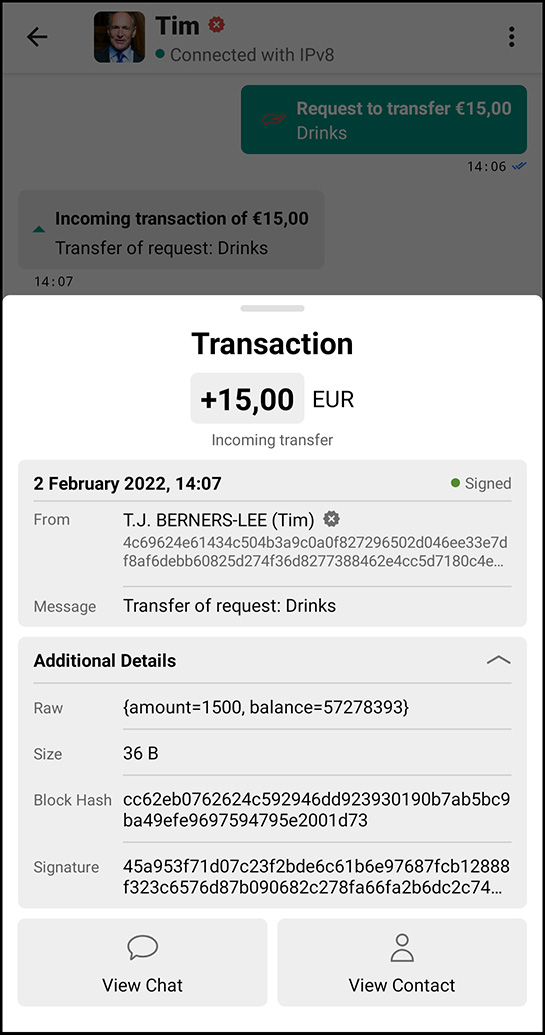} \label{subfig:implementatin_transaction}}
  \caption{Screenshots of implementation}
  \label{fig:implementations}
\end{figure*}

The platform consists of five general views that are interconnected using a navigation bar and direct links. The initial view, the wallet overview, as in \Cref{subfig:implementatin_wallet_overview}, provides a widget-like overview of the identity, exchange, and chats components. For each widget, there is a separate view that includes all related aspects. The information in these widgets is carefully selected to not overflow the user with information.

The base functionality of the designed platform is obviously the societal component. The implementation of a simple chat functionality is present in the form of PeerChat \cite{peerchat-skala2020}, implemented as a community of IPv8. Its core functionality is the exchange of text messages and photos over the P2P network. Many more functionalities had to be implemented to be able to match the market-leading big-tech platforms. The ability to exchange files, locations, contacts, identity attributes, money, and payment requests has been integrated. The chat is an ordered list of the communication between two peers, and every type of message or attachment has its own view. It is again important to not display too much information for the attachments. Detail views display all available information about the attachment. Users have additional access to various convenient chat-related features to improve usability like searching and filtering, muting, archiving, or blocking. To reduce the receipt of unwanted spam the platform discards communication with blocked peers. Large attachments, in particular photos and files, are exchanged using our designed data protocol of \Cref{Subsec:Design-data}. All other attachments are exchanged as single UDP packets and will be delivered using IPv8's default method. If the recipient of a message (or data) is currently not connected, the message cannot be delivered. To track the delivery status of messages, peers are required to acknowledge its receipt. That enables the community to periodically check if the peer is connected and retransmit the unacknowledged messages. In a P2P network it regularly happens that messages can't be delivered as there is no central or distributed component temporarily storing the message. As a consequence, it may happen that two peers are unable to communicate when they are never connected at the same time. 

\begin{figure}[ht!]
\centering
  \subfloat[a][\centering Original PeerChat implementation]{\includegraphics[width=0.5\linewidth]{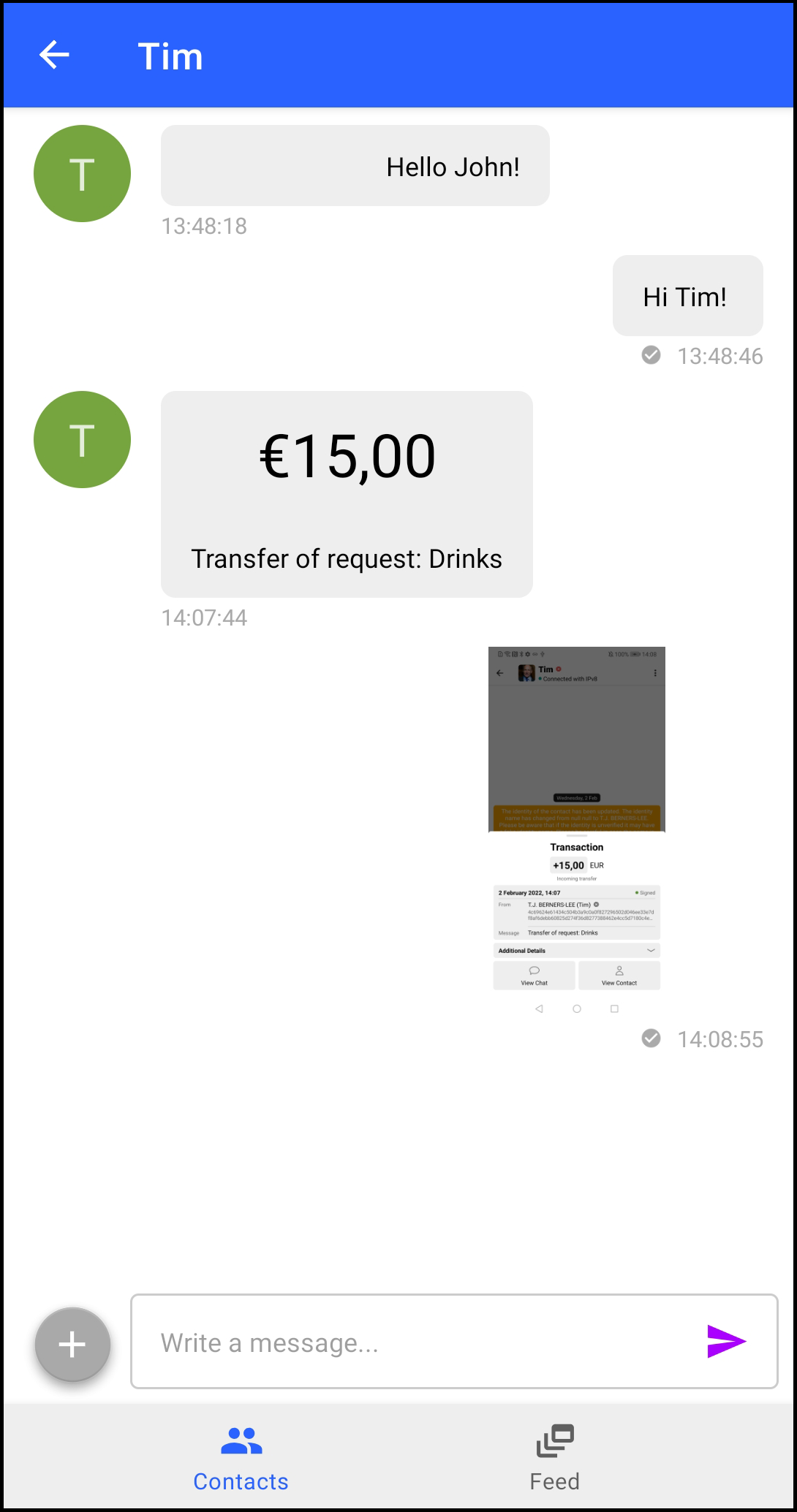} \label{subfig:implementation_chat_old}}
  \subfloat[b][\centering Our chat implementation \cite{timbernerslee}]{\includegraphics[width=0.5\linewidth]{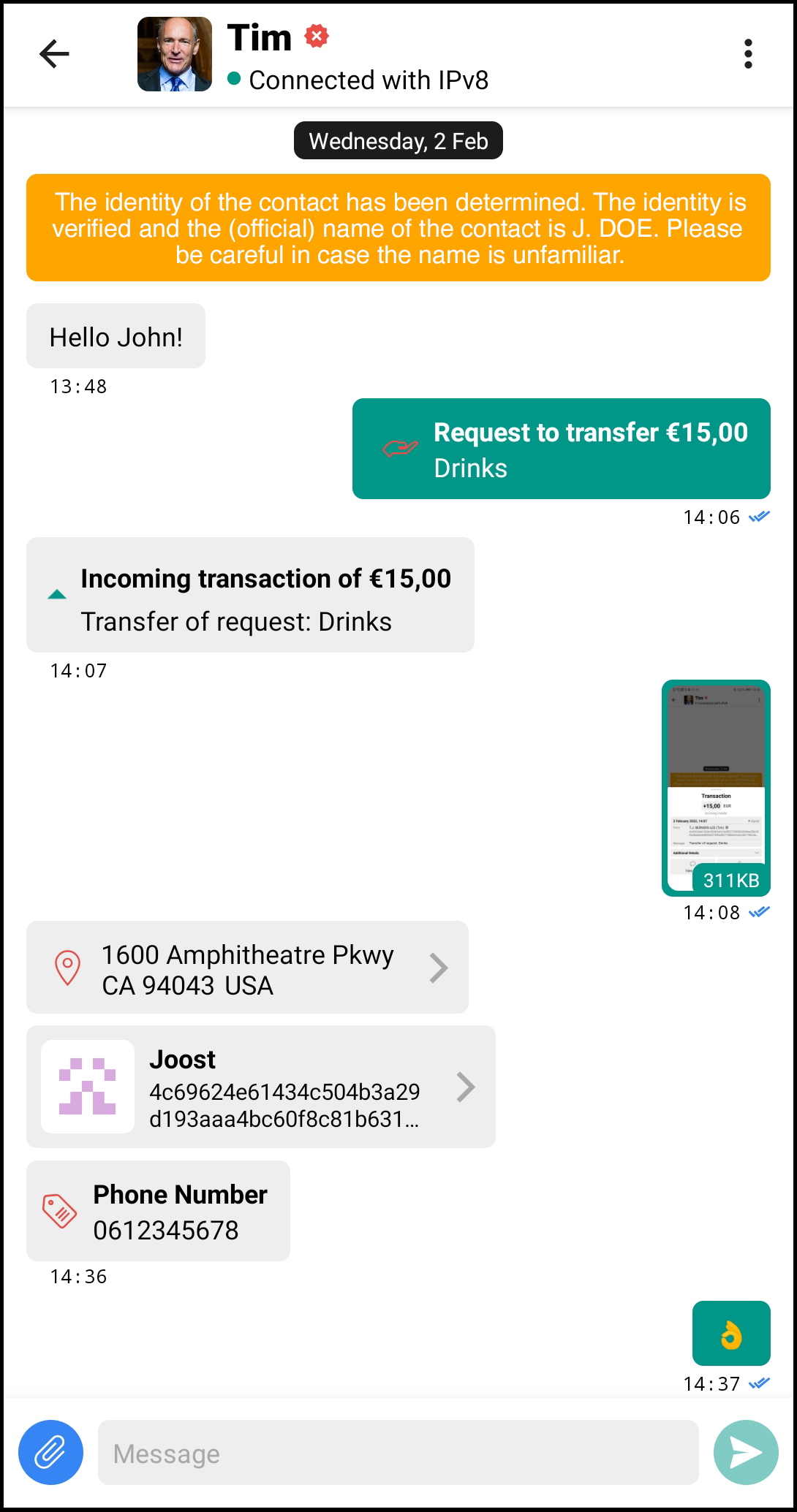} \label{subfig:implementation_chat_new}}
  \caption{Difference between old and new implementation}
  \label{fig:implementation_chat}
\end{figure}

In \Cref{fig:implementation_chat} both the existing PeerChat implementation and our implementation are displayed. Not only is our implementation equipped with much more functionalities, but the design is also concerned about the user experience and ease of use. We've placed options and functionalities behind additional buttons, carefully selected required information to display and provided our platform with full integration and support for QR-codes. The use of QR-codes has many advantages. A lot of information can be embedded in these codes without having to worry about human errors. Inter-platform and offline communication still enable users to exchange the embedded information, even when disconnected from the Internet. These codes can additionally provide direct navigation within the platform, based on the contents of the scanned code. This all combined improves the user experience as less time and effort is required from the user. Example applications of the QR-codes include offline scanning and adding public keys, unspecified payments requests, transfer announcements to and from the exchange portal, and creating and validating identity attestations. Currently, all QR-codes contain unencrypted data as it is not deemed necessary for offline communication. However, future functionalities may for instance include the exchange of information of the self-sovereign identity and would require additional security. QR-codes are perfectly suited to embed encrypted data, but requires the recipient's public key to be known in advance.

\begin{figure}[ht!]
\centering
  \subfloat[a][\centering Identity implementation]{\includegraphics[width=0.5\linewidth]{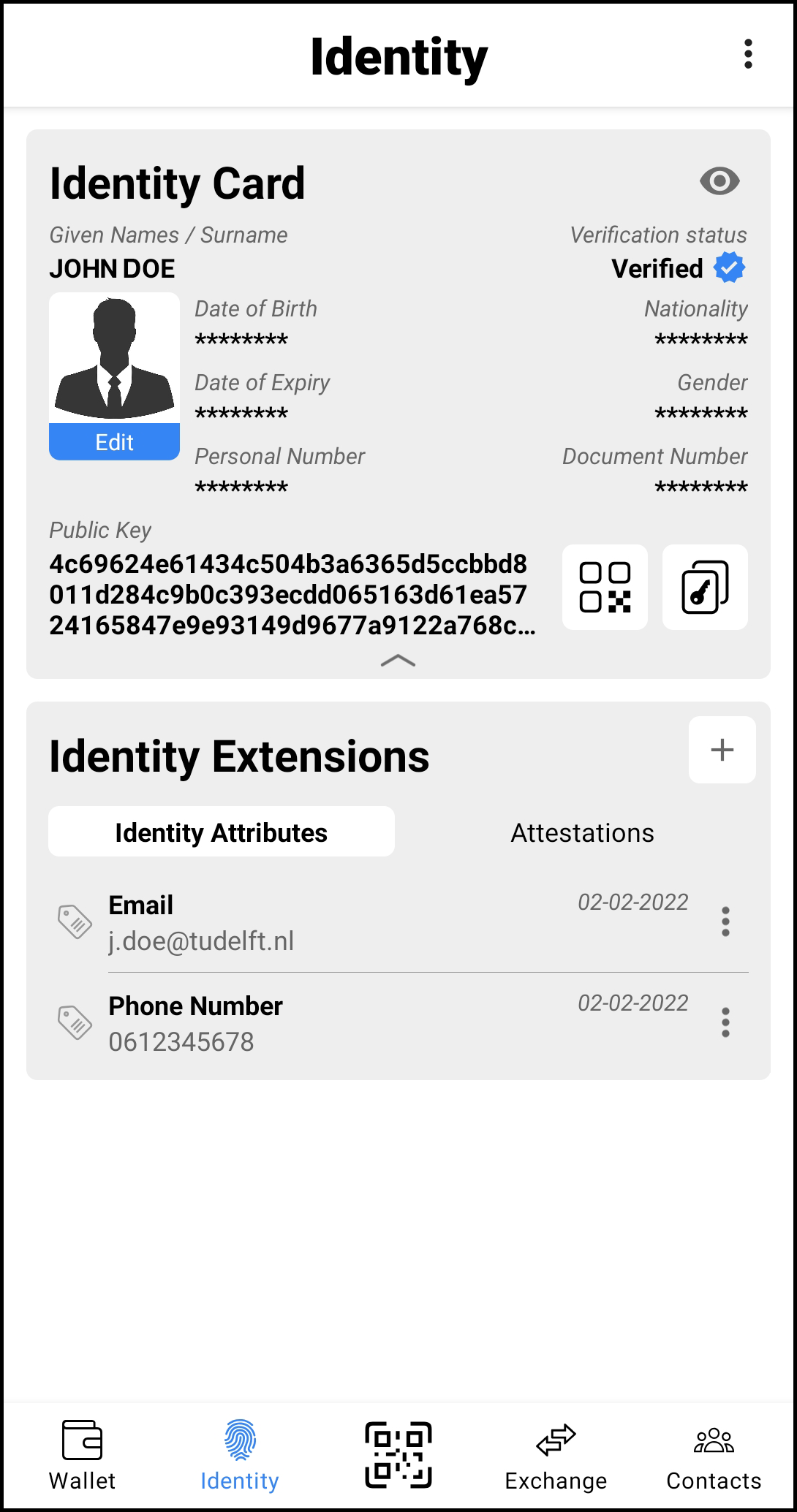} \label{fig:implementation_identity}}
  \subfloat[b][\centering Attestation integration]{\includegraphics[width=0.5\linewidth]{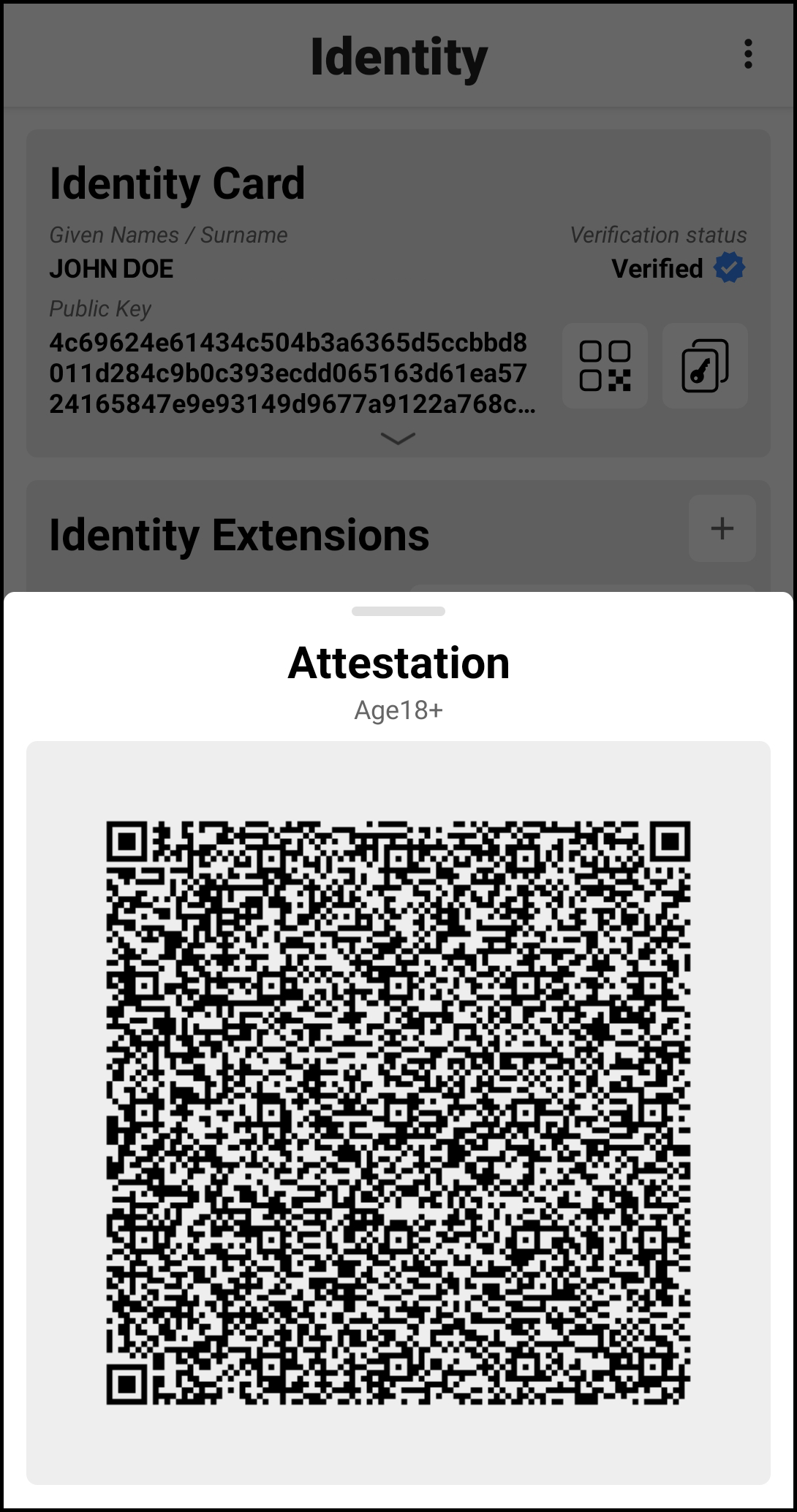} \label{fig:implementation_attestation}}
  \caption{Implementation of self-sovereign identity}
\end{figure}

The integration of the identity component had to be designed from scratch. The identity community handles anything related to the functionalities of the self-sovereign identity, including its storage. Compared to other communities, the identity community does not support communication over the IPv8 network. There is currently no need to additionally share information from the self-sovereign identity, as it may only form a privacy vulnerability. To use the platform, the users are required to onboard their identity as explained in \Cref{Subsec:Design-identity}. Any device with NFC support is required to verify its identity document, as it reduces misuse and provides authenticity. An extra layer of protection makes sure that the identity details are not visible for eavesdroppers, see \Cref{fig:implementation_identity}. Apart from the self-sovereign identity itself, we've also integrated \textit{identity attributes} and \textit{identity attestations}. Identity attributes contain convenient information that serve as an unauthentic extension to the identity. These attributes are for example a phone number, email address, or home address. These attributes are shareable with other peers and solely serve to extend the use of the self-sovereign identity. Validation mechanisms to verify the identity attributes' authenticity are currently not included. The identity attestations are incorporated in unmodified form using the attestation community, as a result of the work of \citet{ssi-rowdy2021}. An example QR-code of an 18+ attestation is portrayed in \Cref{fig:implementation_attestation}. Currently, only the age-related attestation types obtain the age directly from the self-sovereign identity. Future attestation types could embed more confidential information from the identity as it can serve additional purposes. 

\begin{figure}[ht!]
\centering
  \subfloat[a][initial state]{\includegraphics[width=0.9\linewidth]{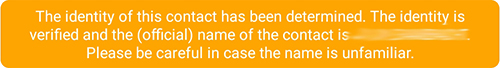} \label{fig:trust-initial}}  \\
  \subfloat[b][updated state]{\includegraphics[width=0.9\linewidth]{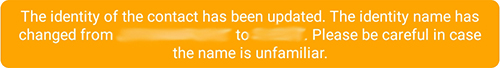} \label{fig:trust-updated}}
  \caption{Detection in alteration of received trust attributes}
  \label{fig:trust_changes}
\end{figure}

The trust attributes, as explained in \Cref{Subsec:Design-trust}, are directly deduced from the self-sovereign identity. These attributes are automatically added to every form of communication with other peers. The user is unable to choose to communicate without sending the trust attributes along, as this damages the integrity of our platform. Changes to the state as a result of the received trust attributes are notified in the chat as in \Cref{fig:trust_changes}. The verification status of the contact is explicitly displayed at several different locations with the sole purpose to draw attention and recognition, especially when unverified. A recognizable blue check star indicates successful verification while a red cross star, that is missing in existing platforms, indicates an unverified peer. Both verification statuses can be seen in \Cref{subfig:implementatin_wallet_overview}.

\begin{figure}[ht!]
    \centering
    \includegraphics[width=0.7\linewidth]{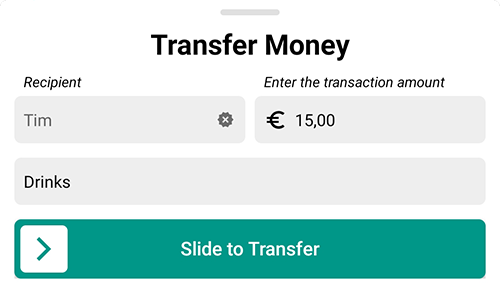}
    \caption{'Slide-to-Transfer' protection}
    \label{fig:implementation_slide_to_transfer}
\end{figure}

An implementation to transfer digital money was already integrated in the \texttt{superapp}, see \Cref{subfig:implementatin_exchange_old}. The eurotoken community handles the wallet and transfer of the CBDC Eurotoken \cite{eurotoken-blokzijl2021} with other peers, while the trustchain community handles the functionalities of the transactions in the distributed ledger. In our platform, the wallet balance is displayed and protected from eavesdroppers by initially hiding the balance, see for example \Cref{subfig:implementatin_wallet_overview}. Various options to exchange money have been integrated as in \Cref{subfig:implementatin_exchange_new}. Firstly, QR-codes are used to deposit or withdraw tokens from and to the exchange portal. Secondly, users have the option to scan a QR-code payment request to transfer tokens, create a direct transfer to another contact, create and send payment requests to contacts over the network or create an unspecified payment request using a QR-code. The latter three are accessible from the chat with the corresponding contact as well. An extra layer of protection, in the form of a 'slide-to-transfer' element as seen in \Cref{fig:implementation_slide_to_transfer}, withholds the accidental exchange of tokens. For convenience, transactions are not only visible in the list of transactions in the exchange view, but also in chats. In \Cref{subfig:implementatin_transaction} a detailed view of a transaction and its contents is displayed. As payment requests are attachments and not formal transactions, they are only included and visible in the chats.

As explained in \Cref{Subsec:Design-data}, our custom data transfer protocol handles the exchange of (large) data blobs. The protocol technically contains an entry point for peers to directly transmit data to another peer, or in some situations to be scheduled. Scheduled transfers are periodically checked and started if the following requirements are satisfied: (I) the peer is connected, (II) there's no other current transfer with the peer, and (III) the size of the transfer does not exceed the maximum allowed size. The protocol exploits packet listeners to be able to directly respond to each of the received packets. The transfer is initiated by announcing the transfer and transfer-specific settings. As (many) other communities may employ the protocol as well, it is convenient to annotate the destined community. As there may be various concurrent transfers with different peers, every transfer must contain a unique identifier to distinguish the transfers from one another. To make sure that both the sender and receiver are in agreement with the transfer parameters, and additionally allow other communities to use different parameters, these parameters are included within the transfer announcement. The transfer announcement and the last data packet of its window must be acknowledged. This acknowledgment either confirms the start of the transfer or receipt of a window of data blocks. In addition to the latter, the current window number and lost packets must be reported. This confirmation consecutively results in the transmission of the next window of blocks until all blocks have been transmitted and received. To be able to stitch all blocks together, the protocol additionally transmits the block number with the packet. The transfer is completed when all blocks have been received. 

To realize a quick response to the transmitted packets, the recurrent packets must contain the minimal required information. The transfer announcement is only transmitted once and is therefore allowed to contain more information. As the data and acknowledgment packets are transmitted numerous times, we've made sure that no redundant information is transmitted. 

\begin{figure}[ht!]
\centering
    \subfloat[a][\centering scheduled status]{\includegraphics[width=0.25\linewidth]{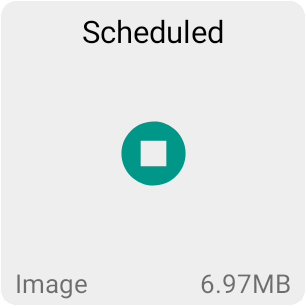} \label{subfig:implementation_download_scheduled}}
    \;\;
    \subfloat[b][\centering progress status]{\includegraphics[width=0.25\linewidth]{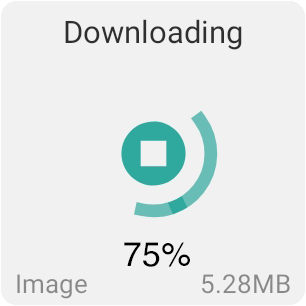} \label{subfig:implementation_download_progress}}
    \;\;
    \subfloat[c][\centering stopped status]{\includegraphics[width=0.25\linewidth]{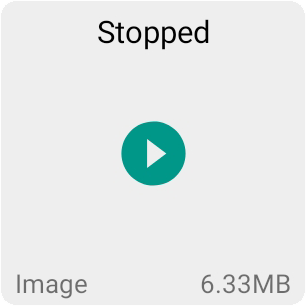} \label{subfig:implementation_download_stopped}}
    \caption{Download progress indicators in chat}
    \label{fig:implementations_download_indicator}
\end{figure}

To accommodate communities and users with convenient information during and after the transfer, the protocol has built-in support for callbacks. These callbacks enable the execution of specific tasks after the transfer progressed to another state. These tasks differ for each application and are therefore outside the scope of the protocol. The sender is able to execute specific code after the data has been successfully sent or upon receipt of an error. The receiver has access to the transfer progress, transfer completion, and erroneous updates. Specifically, the former two are important to our platform. Transfer progress updates enable the application to display the current download status to the user, see \Cref{fig:implementations_download_indicator}. For convenience in certain situations, the user is able to stop and restart the transfer at a later time. The transfer complete update triggers the conversion of the raw binary data to the correct format, creates an external file on the phone storage, and visually embeds it in the chat. The protocol is dependent on many parameters that may have an impact on the performance. In \Cref{Sec:ExperimentalAnalysis} these parameters are analyzed to obtain the optimal performance during \textit{normal} operation.

\begin{figure}[ht!]
    \centering
    \subfloat[][Unit tests]{\includegraphics[width=0.8\linewidth]{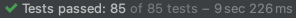}} \\
    \subfloat[][Code coverage]{\includegraphics[width=0.45\linewidth]{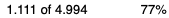}}
    \caption{Code quality}
    \label{fig:code_quality}
\end{figure}

We've additionally analyzed the code quality of the protocol. A high code coverage is required to reduce bugs and ensure correctness. The protocol, including all associated methods and classes, is tested using 85 unit-tests that cover about 77\% of the code (\Cref{fig:code_quality}). 
\section{Experimental Analysis and Evaluation} \label{Sec:ExperimentalAnalysis}
In the previous sections, we have discussed the design and implementation of our data transfer protocol. In this section, an experimental analysis is performed to derive the optimal protocol settings \underline{under normal operation}. We've only considered normal cases in this analysis due to missing theoretical and applied expertise. We also performed an evaluation of a large-sized transfer with these optimal settings to prove its contribution and applicability to our platform.

\subsection{Experimental Analysis}
To exploit the best possible performance, the designed binary data transfer protocol requires its settings to be optimal. We define the protocol to be optimal if (I) the transfer speed is as high as possible, (II) the number of lost packets/blocks (as explained in \Cref{Subsec:Design-data}) is as low as possible, and (III) the number of retransmitted windows of blocks (by the sender) and acknowledgments (by the receiver) is as low as possible. The second constraint does not necessarily contribute to higher transfer speeds as lost packets are embedded in the next window of blocks. The last constraint should contribute to higher transfer speeds because no windows of blocks have to be retransmitted and there's no additional idle time waiting for an acknowledgment. The analysis is performed in ideal situations and focuses on the general picture, i.e. aspects as delays due to latency and packet loss are not included.

As mentioned before, the UDP packet size is limited due to Ethernet constraints. As two packets with a payload of 500 bytes carry twice as much redundant information as one packet of 1000 bytes, a transfer using a greater block size $B$ is preferred and should theoretically have a positive impact on the runtime. The maximum \textit{data} size of UDP packets for IPv8 has been determined (through trial-and-error) to be around 1241 bytes. The exact size may depend on each peer, the chosen packet header options (encryption, signature, public key, etc.), and the data packet metadata. To keep a safe margin we've decided to allow data of at most 1200 bytes in each packet. The window size $W$ is defined as the number of bytes ($n_{\text{blocks}} \times$ block size) the sender can transmit without having to wait for an acknowledgment of receipt from the receiver. Theoretically, a greater window size would directly contribute to larger \textit{transfer speeds} as there are less windows of blocks to be transmitted. Also, a smaller number of acknowledgments has to be sent and received, reducing the overall idle time of both participants. Greater window sizes also increase the existence of late or lost blocks, specifically in imperfect or congested networks, with an increased number of retransmissions of windows of blocks as a result. The importance of this analysis is to find the best trade-off between a great window size and low delay due to lost packets.

The other parameters do not directly impact the performance, apart from the block and window size. The retransmit interval may affect the performance when it is either too tight or loose, but it will only play a role in a small part of the cases. A tight interval can force windows of blocks or acknowledgments to be retransmitted while they are still in transit and may arrive shortly after. For a loosely set interval, the protocol may unnecessarily have to wait for a window or acknowledgment. The transfer timeout interval is less critical and will only affect the performance when a window or acknowledgment has abused all retransmit attempts. The retransmit attempt count likewise has little influence on the performance. \\

\subsubsection*{Experimental Setup}
The experimental setup is equipped with two phones, a Xiaomi Redmi 9T with Android 10 and a Huawei P20 Lite with Android 9, both 4GB RAM. The phones have installed the same version of the app and are connected to the same WiFi-6 mesh network (NETGEAR Orbi RBK753). To obtain more accurate results, each experiment is executed five times. Also, to verify the independence of the file size on the transfer, the experiment is executed for multiple file sizes. Each important step of the protocol is captured in a log to be processed in Python. An automatic Kotlin script makes sure that every combination of parameters and the five iterations are executed consecutively. \Cref{table:exp_analysis_parameters} gives an overview of the analyzed parameter values. The maximum parameter settings are windows of 128 blocks of 1200 bytes each, equivalent to 150kB of unacknowledged data that is the fundamental limit to the data transfer performance. With a certain unknown latency this results in an exact transfer speed limit. As no a packet loss and latency emulation has been performed, we cannot determine this theoretical limit.

\begin{table}[ht!]
	\centering
	\caption{Parameters that are being tested for optimal execution. The number of iterations have only been used for consistency. In total 56 combinations of parameters have been executed 5 times.}
	\begin{tabular}{ l l l }
		\textbf{Parameter} & \textbf{Values} &	\\ 
		\hline
		Block size ($B$) & $600, 700, 800, 900, 1000, 1100, 1200$ & [bytes] \\
		Window size ($W$) & $16, 32, 48, 64, 80, 96, 112, 128$ & [blocks] \\
		\hline
		Iteration & 0, 1, 2, 3, 4 & [-] \\
		\hline
	\end{tabular}
	\label{table:exp_analysis_parameters}
\end{table}

\begin{figure*}[t!]
	\subfloat[][]{
		\label{subfig:transfer_speed_10}
		\includegraphics[width=0.48\textwidth]{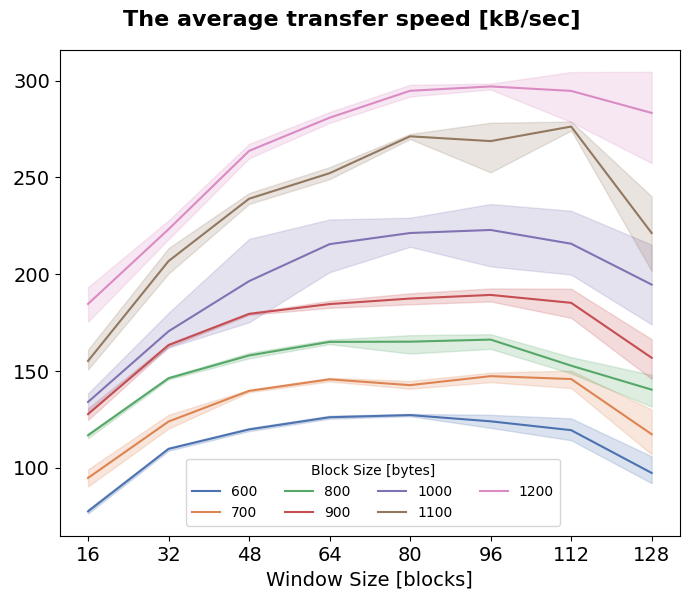}
	}
	\subfloat[][]{
	\label{subfig:packets_loss_10}
		\includegraphics[width=0.48\textwidth]{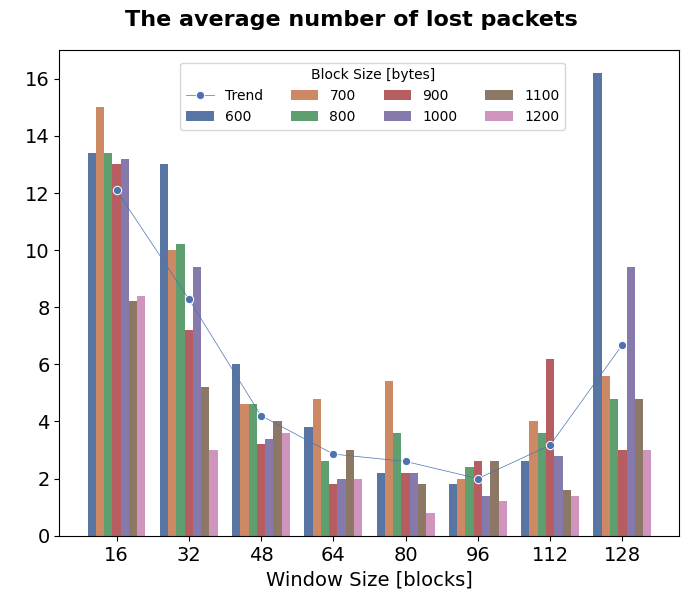}
	}
	\\
	\subfloat[][]{
	\label{subfig:retransmits_window_10}
		\includegraphics[width=0.48\textwidth]{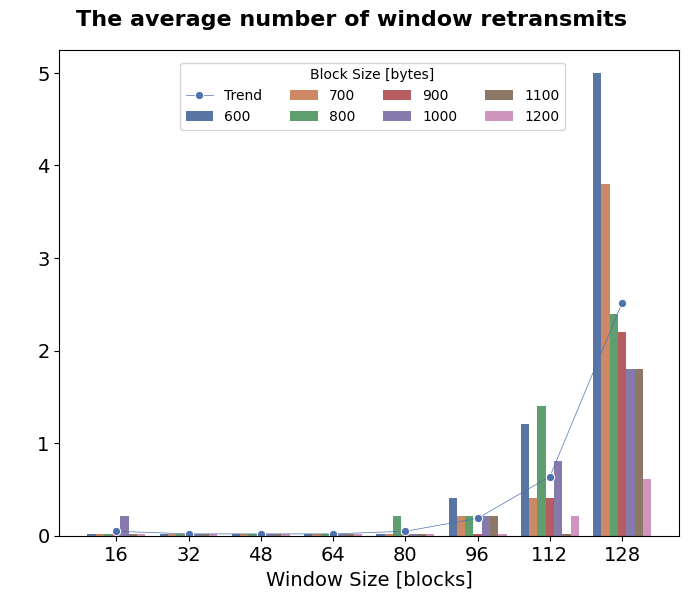}
	}
	\subfloat[][]{
	\label{subfig:retransmits_ack_10}
		\includegraphics[width=0.48\textwidth]{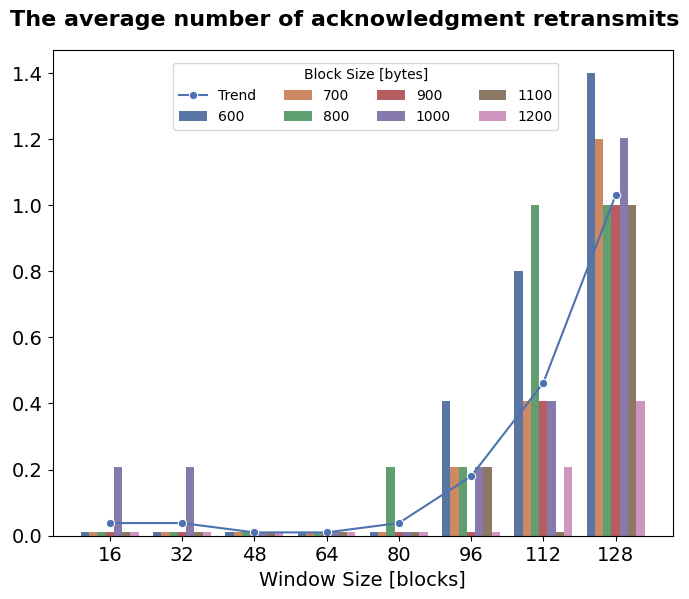}
	}
	\caption{The results of all executed tests for each variable window and block size, averaged over five iterations.}
\end{figure*}

\subsubsection*{Experimental Results} 
The optimality of the performance of the protocol can be determined in combination with the before mentioned requirements. \\
The results of the first requirement, the transfer speed of the protocol, is displayed in \Cref{subfig:transfer_speed_10}. We can clearly see the effect of the variation of the block and window size. Higher block sizes increases the transfer speed. The window size follows a parabolic curve and the transfer speed is optimal for a window size $W = 80$ and $W = 96$ blocks. We cannot yet determine the optimal value for the window size as these values are very similar and are less pronounced than the block size. To decide on the optimal sizes, we have to include the requirements as well.

The results for the second requirement, the number of lost packets/blocks during a transfer, is visualized in the plots of \Cref{subfig:packets_loss_10}. The block size shows a slightly decreasing pattern and overall contains the lowest number of lost packets for greater block sizes. There is no consensus on the block size \textit{within} each window size, as there are small deviations and not consistently decreasing or increasing. The trendline, a combined average of all block sizes within each window, shows a minimum for the same window sizes as of the first requirement, but slightly favors $W = 96$ blocks. For window sizes greater than $W = 96$ blocks the number of lost packets again increases. The protocol is behaving more unreliable as more lost blocks have to be added to the next window. Every lost block will cause a marginal decrease of the transfer speed as more blocks have to be delivered in the next window. The aim remains to reduce these lost blocks as much as possible. Especially for more unreliable connections, the impact of lost blocks may become more pronounced as the transfer progresses further. 

The last requirement, the number of retransmitted windows of blocks and the number of retransmitted acknowledgments, are entangled as it takes both the sender and receiver in the equation for the same transfers. In \Cref{subfig:retransmits_window_10} and \ref{subfig:retransmits_ack_10} the results for the sender and receiver are visualized, respectively. Both diagrams show the same pattern. The optimal block size again shows no notable preference within each window size. The results for the window sizes are a strong indicator that we don't want them to be oversized. The number of retransmitted windows and acknowledgments is of neglectable proportion for a window size less than $W = 112$ blocks, or even $W = 96$ blocks if we would be really strict. As the protocol has to wait for a full interval for every retransmission, it has a big impact on the overall performance. It is crucial to reduce the number of retransmissions to an absolute minimum. For the largest three window sizes, a small experiment was executed that verified if the large increase of retransmissions were the result of a too tightly set retransmit interval. The interval was increased majorly, just for verification purposes, and solely served for additional analysis of the third requirement, without including the results of the other requirements. The results showed that the number of retransmissions slightly improved, but the pattern was equally in place. It was deemed unnecessary to investigate it further.

\begin{figure*}[ht!]
	\centering
	\includegraphics[width=0.67\linewidth]{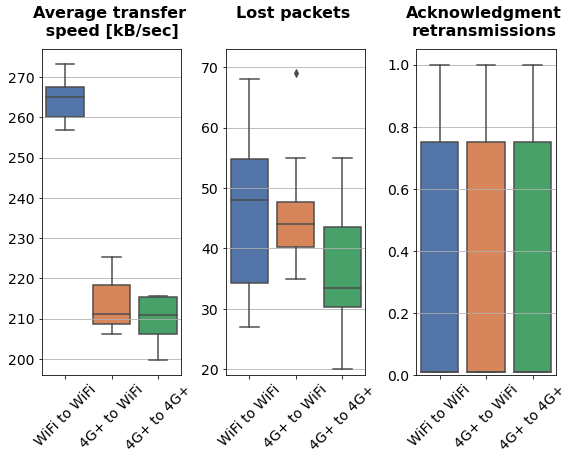}
	\caption{Evaluation of the performance of a transfer of 250MB using the optimal window size of $96$kB, executed 10 times.}
	\label{fig:evaluation_performance}
\end{figure*}

We must take into account that the experiments have been performed under somewhat optimal circumstances: phones only running system services and the platform itself, and both connected to the same \textit{local} WiFi network. From this, we could argue that for worsened conditions, the transfer speed and the number of lost packets and retransmissions would logically increase. We can combine and summarize our findings based on the results of the experiments and analysis. The performance of the transfer speed (continuously) increases with greater block sizes, with the optimal block size of $B = 1200$ bytes. For the other requirements, the difference in block size was less definite. The transfer speed found an inconclusive optimal for window sizes of $W = \{80, 96\}$ blocks. The second requirement showed similar results, although slightly in favor of the latter. Both types of retransmissions for the third requirement indicated an increase of retransmissions for greater window sizes, specifically above $W = 80$ blocks. To obtain the optimal performance, in combination with the minimal number of lost blocks and retransmissions, we can conclude that a window size of $W = 80$ blocks is optimal under normal operation. The optimal window size based on our findings is thus $W = 96000$ bytes or $96$kB. We've deduced these optimal parameters for ideal situations, and therefore does not include lateness and loss of blocks. 

We've additionally executed the same experiments using multiple file sizes to verify its independence on the performance. All results are similar and seem to indicate that the file size is independent on the performance. We've also concluded that an increase of the retransmit interval does not necessarily give a significant overall reduction of the number of retransmissions, and therefore not contributes to better performance. 

\subsection{Performance Evaluation}
Now that we've determined the optimal parameters for the protocol, we want to see how it performs in the wild for a large-sized transfer of 250MB. This enables us to evaluate the performance more consistently over a longer period of (transfer) time. In most cases, phones are not connected to the same local WiFi network. We have to consider three commonly-used situations (I) WiFi to WiFi, (II) WiFi to 4G+, and (III) 4G+ to 4G+. For the connections of 4G+, we use the telecom providers Vodafone and KPN on the same phones as before. Our evaluation includes the same aspects as the experimental analysis. Instead of finding the optimal parameters, we this time evaluate the applicability and the difference between the three situations. Each experiment is executed 10 times to obtain more consistent results. We expect the first situation to offer the highest transfer speed and least number of lost packets and retransmissions as the packets are only exchanged within the local network. 

The performance results for each of the connection types are portrayed in \Cref{fig:evaluation_performance}. The transfer speed shows a clear division in performance, in favor of the inter-WiFi transfer of about 260kB/s. The transfer speeds for the two situations using a mobile connection have very similar speeds of about 213kB/s and 210kB/s. It's a good indication that there is no extreme performance difference between an exchange using one WiFi-connected device and a complete mobile network exchange. The transfer speed of the inter-WiFi transfer is executed on a local network and therefore sketches a slightly biased image. The transfer speed of an exchange between two non-local WiFi networks would theoretically be lower, and possibly be more similar to the other situations. However, we do notice a bigger spread in terms of lower speeds for the complete mobile network exchange. If we look at the absolute speed, we must conclude that our protocol is nowhere near the download speeds of current network infrastructures. The exact reason is unknown, but it is expected to be some limitation in IPv8 and possibly sub-optimal UDP socket buffering.

The number of lost packets contradicts our expectation. The number of lost packets is on average larger for an inter-WiFi transfer in comparison with the other two situations. In absolute numbers though, we can conclude that the total loss of about 50 packets in a transfer of over 200.000 packets, equivalent to one lost packet within every fifty windows, is neglectable. This number would be much higher for sub-optimal connections. No retransmission of windows of blocks were encountered, and therefore left out of the figure. The number of retransmissions of acknowledgments are very equivalent for all three situations, and only allows one acknowledgment retransmission per transfer on average as result of unresponsiveness. 

From these results we can conclude that our protocol has been performing at the top of its \textit{abilities} in normal operation. The absolute transfer speed is disappointing. During the experiments, it was noticed that larger-sized transfers experienced memory allocation issues on the phones. Currently, the protocol stores the sent and received data in memory for reconstruction purposes. The current maximum file size has therefore been limited to 250MB, but can differ from phone to phone.
\section{Time Management}
The research, design, implementation, analysis, and documentation have been an effort of one person in roughly nine full-time months. The research phase required about two months to study the literature and existing platforms, including the basics and characteristics of IPv8, TrustChain, and other components. The design and implementation phase were entangled as the scope of the research widened several times along the way. You could say that it was a repeating process of invent-design-implement for most of the features. An initial layout of the application was designed and implemented to provide a basic platform that slowly evolved to its final state. This included familiarizing the style and coding within the \texttt{Trustchain superapp} and the \textit{Kotlin} language. The design and implementation of features were not only concerned with the features themself, but also the UX and UI design of the platform. In total, about six months have been allocated to the design, implementation, and analysis of the platform. The data transfer protocol is the only component that has additionally been analyzed as part of experimental analysis and evaluation. Of these six months, about one month of work was required to optimize and analyze the data transfer protocol. The last month, and also some time earlier in the process, was dedicated to documenting and finalizing this paper.
\section{Conclusion}
Ownership and exchange of sensitive or private information has never been a more relevant topic, also due to the COVID pandemic. This paper has presented a novel Web3 platform for identity, trust, money, and data. Decentralization partially solves the lack of self-sovereignty for identity, money, and data in the current online world. We performed the first exploratory study that shows the viability of the integration of a self-sovereign identity in a social platform in a useful, secure, and private fashion. The application of self-sovereign identities has additionally shown to be effective in providing trust to peers in the P2P network. Governments may change their mind about their identity management systems in the near future. Many centralized tasks can be replaced, providing their citizens more power and ownership over their data, while retaining authenticity and majorly reducing costs. Central banks have started designing Central Bank Digital Currencies globally. Our platform incorporates the exchange of digital money in a private and informal way, similar to what cash once was. Data and personal information has been owned and managed by big-tech companies for far too long. The centralized structure is the core of the problem in online communication. Our P2P data transfer protocol enables peers to securely and privately exchange messages and data while reducing the leakage of metadata to a minimum. 

Unfortunately, we cannot neglect some major disadvantages in our platform as well. Firstly, the availability and connectivity of peers remain an open issue as it introduces lateness in the delivery of messages and data. People are currently not used to lateness in existing platforms which use central servers or nodes. Secondly, peers in a fully P2P network must keep themselves online by constantly discovering and connecting to peers with the consequence of draining the phone's battery. Improvements could be made, but it will always remain a weak spot of P2P systems. Finally, the implementation of the CBDC EuroToken is dependent on central components and not yet reliable enough. 

Overall, we conclude that we've designed a well-functioning platform that incorporates all aspects of our research in a valuable, private, and secure manner. There is still a lot to discover and we're curious to see what direction big-tech companies, governments, and banks will pursue in the near future.

\bibliography{bib}

\end{document}